# Nonadiabatic kinetics in the intermediate coupling regime: comparing molecular dynamics to an energy grained master equation

Darya Shchepanovska[1], Robin J. Shannon[1], Basile F. E. Curchod[2],*, and David R. Glowacki[1,3,4],*

[1]School of Chemistry, University of Bristol, Bristol, BS8 1TS, UK; [2]Department of Chemistry, Durham University, Durham, DH1 3LE, UK; [3]Intangible Realities Laboratory, University of Bristol, BS8 1UB, UK; [4]Department of Computer Science, University of Bristol, BS8 1UB, UK

*glowacki@bristol.ac.uk; basile.f.curchod@durham.ac.uk

## Abstract

Here we outline and test an extension of the energy grained master equation (EGME) for treating nonadiabatic (NA) hopping between different potential energy surfaces, which enables us to model the competition between stepwise collisional relaxation and kinetic processes which transfer population between different potential energy surfaces of the same spin symmetry. By incorporating Zhu-Nakamura theory into the EGME, we are able to treat nonadiabatic passages beyond the simple Landau-Zener approximation, along with corresponding treatments of zero-point energy and tunnelling probability. To evaluate this NA-EGME approach, we carried out detailed studies of the UV photodynamics of the volatile organic compound $C_6$-hydroperoxyaldehyde ($C_6$-HPALD) using on-the-fly ab initio molecular dynamics and trajectory surface hopping. For this multi-chromophore molecule, we show that the EGME is able to quantitatively capture important aspects of the dynamics, including kinetic timescales, and diabatic trapping. Such an approach provides a promising and efficient strategy for treating the long-time dynamics of photo-excited molecules in regimes which are difficult to capture using atomistic on-the-fly molecular dynamics.

## 1. Introduction

The accuracy of molecular photodynamic simulations in the excited state is inherently constrained by the dimensionality of the system. Exact non-relativistic quantum mechanical dynamics of a wavepacket can be described by solving the time-dependent Schrödinger equation, but exponential scaling limits this approach to small molecular systems. At the opposite end of the scale, there is a growing interest in describing the nonadiabatic dynamics of very large systems characterised by exciton transfer between chromophores.[1] In fact, an analytical description of nonadiabatic transitions for a simple one-dimensional two-state system in the weak coupling limit has been available since 1932, developed simultaneously, and separately, by Landau, Zener, and Stueckelberg.[2] In many cases Landau-Zener (LZ) theory works reasonably well even for larger, multidimensional systems. Later, Zhu and Nakamura built on this framework to produce a set of exact nonadiabatic transition probabilities for different types of nonadiabatic curve crossings.[3] Zhu-Nakamura (ZN) theory is valid over the entire coupling regime, is fully analytical, and incorporates tunnelling contributions. Like LZ theory, it is formulated in a single dimension.

On-the-fly trajectory-based semiclassical dynamics accounts for the full dimensionality of a molecular system. For example, Tully's fewest switches surface hopping (FSSH) is a well-known and efficient way of simulating femtosecond timescale processes in the excited state,[4] where the time evolution of a wavepacket is approximated by a swarm of independent trajectories that classically propagate the nuclear degrees of freedom on a potential energy surface (PES) calculated on-the-fly.

Each trajectory can stochastically switch between electronic states in regions of strong coupling. While FSSH has known shortcomings (including overcoherence, and neglect of tunnelling and interference effects[5]), it often provides an accurate and scalable method that is now widely used to explore photodynamic phenomena, also for atmospheric chemistry.[6] Given that FSSH typically has a sub-femtosecond integration time step, pushing the simulation into the nanosecond regime necessitates compromises with respect to the electronic structure method and number of trajectories. It has been suggested that this bottleneck might be overcome through machine learned energies and couplings.[7]

For longer timescale simulations in the statistical regime, alternatives to conventional nonadiabatic dynamics strategies are needed. The energy grained master equation (EGME) is the numerical implementation of the exact master equation which discretises the density of states $\rho$, recasting it in matrix form. The EGME has recently been applied to the study of non-RRKM reaction kinetics in the gas phase,[8] in solution,[9] and in surface chemistry.[10] Unlike FSSH, where an electronic structure calculation is performed at each step of a trajectory, an EGME calculation needs only the energies, frequencies and rotational constants of the relevant stationary points. This allows for the use of more computationally demanding electronic structure calculations and detailed sensitivity analyses on the results. The EGME also enables treatment of collisional activation and dissipation from the system. Furthermore, unlike molecular dynamics simulations where zero-point energy can leak, vibrational zero-point energy at the stationary points can be included explicitly in an ME calculation. Approaches to zero-point energy conservation in quantum-classical trajectories exist,[11] but they are not adopted in the standard fewest switches surface hopping (FSSH) algorithm used in this work. Tunnelling corrections may also be included in the framework of the EGME approach using an asymmetric Eckart barrier[12] or semiclassical WKB theory.[13] Solving the ME returns temperature and pressure dependent species profiles, making it a useful tool for modelling atmospheric or interstellar reactions.

Nonadiabatic analogues of standard statistical rate theories generalise classical transition state theory (TST) to reactions involving multiple PESs.[14] For example, intersystem crossings have been successfully modelled by using both the LZ and ZN expression for the inter-state surface hopping probability at the minimum energy crossing point (MECP).[15] Until recently, simulation of intersystem crossings in the surface hopping framework has been limited by the need for the global calculation of the spin-orbit coupling matrix elements[16] or spin-orbit coupling gradients.[17] It is now possible to run semiclassical trajectory surface hopping simulations that include coupling between states of arbitrary multiplicity.[18] Using the LZ approach to describe coupling between states with differing multiplicity works well in the weak coupling regime, but fails for strong coupling.[2a] A nonadiabatic EGME model (NA-EGME) of internal conversion should instead use the ZN expression, which is able to accurately treat the analytical nonadiabatic transition probabilities for the full range of energies and couplings,[3b] giving the LZ result in the weak coupling limit, and the transition state theory result in the strong coupling limit. The ZN description of the coupling region can also be formulated to include contributions from tunnelling through the crossing barrier. In contrast to the full-DOF description of quantum-classical dynamics, the ZN equations are only formulated for 1-D crossings. Herein, we provide evidence that – for seam-like crossings – the ZN approach offers a good approximation to describing nonadiabatic transitions between adiabatic states.

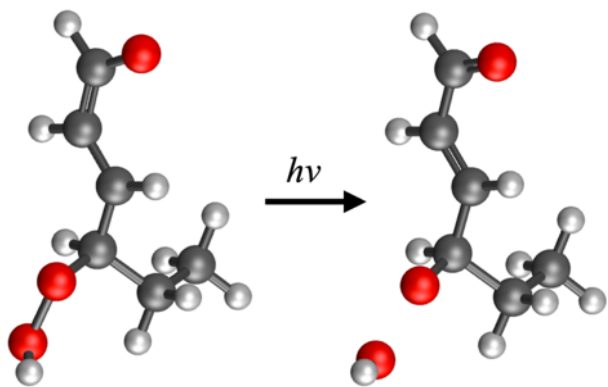

Figure 1: Main reaction channel for $C_6$-HPALD photodissociation in 300-400 nm range.[19]

In this work we apply the NA-EGME to predict the dissociation rate of a photoexcited bi-chromophoric hydroperoxy aldehyde, $C_6$-HPALD, whose primary photodissociation channel is shown in figure 1. In order to evaluate the validity of this approach, we show that the NA-EGME results are similar to the outcomes of a FSSH molecular dynamics simulations. HPALDs represent a class of molecules important in atmospheric chemistry, and it has been suggested that they participate in OH radical recycling at low $NO_x$ regions of the troposphere.[20] As a product of isoprene oxidation, they contain an α,β-enone chromophore which absorbs sunlight in the UV range, which is in close proximity to a labile peroxide bond. Previous experiments investigating the photodissociation kinetics of $C_6$-HPALD reported OH radical production under UV light.[19] In this paper, we are less concerned with the atmospheric details of HPALD photodissociation, but rather with $C_6$-HPALD as a prototypical example of a multi-chromophore system with an interesting seam-like nonadiabatic coupling topology between two low-lying excited states. We show that the dissociation rate obtained using a NA-EGME model is in agreement with the results of nonadiabatic FSSH dynamics, at a significantly reduced computational cost.

This paper will be structured as follows. Firstly, we describe how nonadiabatic effects can be included in an EGME model using ZN transition probabilities. Second, we describe the electronic structure calculations used to characterise the relevant excited states of $C_6$-HPALD in the Franck-Condon region, and along the dissociative coordinate. Third, we describe the FSSH and NA-EGME calculations and their adiabatic counterparts. Fourth, we compare the results of these contrasting methods for a single isolated reaction coordinate which corresponds to the wavepacket moving over a transition state on the $S_1$ surface leading to photodissociation. Lastly, we extend both models to include all rotational conformers of $C_6$-HPALD.

## 2. Methods

### 2.1 Constructing a nonadiabatic master equation

The energy grained master equation (EGME) is a Markov-state model that has found widespread application to non-equilibrium problems in chemical kinetics.[9, 21] Most applications of EGME models focus on reactive processes on a single electronic state and do not incorporate nonadiabatic coupling. While strategies for calculating microcanonical rate coefficients for nonadiabatic processes do exist, to our knowledge there has only been one attempt to incorporate such transitions into an EGME framework: specifically, Plane *et al.*[21b] modelled temperature and pressure dependent intersystem crossing kinetics by treating the extended seam between the singlet and triplet state as a dividing surface. The minimum energy crossing point (MECP) between the states was treated as a pseudo-transition state, and the probability of spin-forbidden hopping transitions between these states was calculated by convoluting the density of states at the MECP with the LZ expressions to obtain microcanonical rate coefficients.

Building on the work of Plane *et al.*, we have extended the NA-EGME to calculate the rate of HPALD photodissociation, where the nonadiabatic transition of interest involves coupling between two states of the same multiplicity – i.e., significantly stronger coupling than the intersystem crossing investigated by Plane et al. The LZ model is ill-suited to internal conversion as it assumes the inter-state coupling is localised, and weak. For these reasons, we used ZN theory to describe nonadiabatic transition probabilities in the coupling region. The ZN equations produce the correct analytical hopping probability coefficients over the full range of coupling regimes, for a 1-D nonadiabatic tunnelling type crossing,[3b, 22] returning the LZ result in the limit of weak coupling, and the classical transition state theory result in the limit of strong coupling.

An adiabatic master equation model (A-EGME, illustrated in figure 2) is constructed from any number of connected potential energy wells (isomers) and the transition states between them. In order to make the problem computationally tractable, the energy of each species is discretised into bins or grains of a set size. The population density across each energy grain of every isomer in the system is then defined by a vector **n**($E,t$), and it is possible to formulate a set of coupled differential equations in terms of **n**($E,t$) that describe the time-evolution of the grain populations. Recasting these differential equations in matrix form defines the chemical master equation.

$$\frac{\partial \mathbf{n}(E,t)}{\partial t} = \mathbf{M}\mathbf{n}(E,t) \qquad (1)$$

The matrix **M** is expressed as [ω(**P**-**I**)-k], where ω is the Lennard Jones collision frequency, **P** is a matrix of transition probabilities between grains, **I** is the identity, and k is a diagonal matrix of energy-resolved microcanonical rate constants, $k(E)$, for the reactive process. In the EGME, population transfer between grains can arise due to interactions with a bath or through reactive loss/gain to a connected isomer. Energy transfer as a result of bath interactions is typically modelled using an exponential down model. For reactions between different isomers, population transfer can only occur between corresponding grains of the same energy. This is included in the model through unimolecular microcanonical rate coefficients $k(E)$ calculated from Rice-Ramsperger-Kassel-Marcus (RRKM) theory. The RRKM microcanonical rate coefficient at energy $E$ is

$$k(E) = \frac{W(E)}{h\rho(E)} \qquad (2)$$

where $W(E)$ is the sum of rovibrational states at the optimised transition state geometry, and $\rho(E)$ is the density of rovibrational states at the isomer.

Because the EGME can be used to model out-of-equilibrium phenomena,[23] it is often applied to reactions in atmospheric and combustion chemistry which cannot be modelled with equilibrium TST techniques due to the non-Boltzmann distribution of energy in isomers. This allows us to replicate the non-equilibrium energy distribution of a wavepacket directly following photoexcitation.

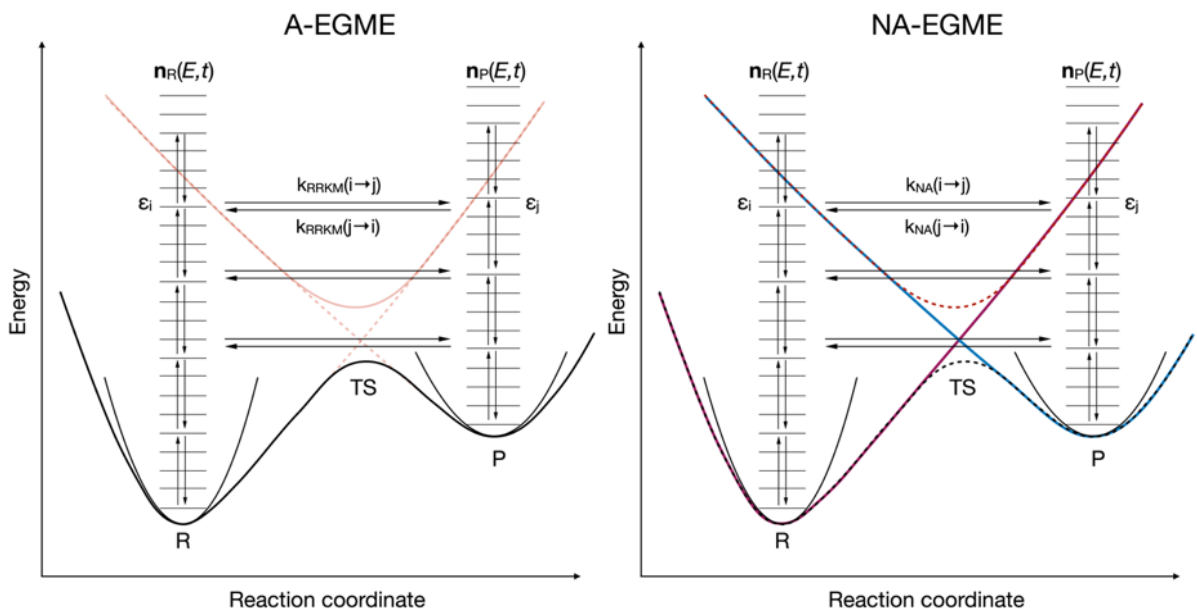

Figure 2: Left hand panel shows a standard A-EGME model describing a reactant and product species connected by a transition state. Right hand panel shows the NA-EGME, with nonadiabatic transitions integrated through the inclusion of energy resolved ZN transition probabilities to the upper state.

The NA-EGME model, illustrated in figure 2, is constructed analogously to the conventional ground state A-EGME, except that the microcanonical rate coefficients are not calculated through RRKM theory. Nonadiabatic coupling between states is included through an energy resolved ZN expression for the transition probabilities $P_{ZN}$ between two diabatic states in the vicinity of a crossing point. We can then compute a set of nonadiabatic microcanonical rate coefficients $k_{NA}(E)$ which transfer population between the different diabatic states. The expression for $k_{NA}(E)$ is similar to the RRKM expression in equation 2,

$$k_{NA}(E) = \frac{N_{TS}(E)}{h\rho_{S_1}(E)} \quad (3)$$

where the density of states at the optimised $S_1$ minimum is $\rho_{S_1}$ and $N_{TS}$ is the convolution of the ZN transition probabilities $P_{ZN}$ and density of states at the transition state $\rho_{TS}$.

$$N_{TS}(E) = \int_0^{E'} \rho_{TS}(E - E')P_{ZN}(E)dE' \quad (4)$$

The complete equations for $P_{ZN}$ are available in the SI, and implemented in MESMER (master equation solver for multi-well energy reactions).[24]

## 3. Computational details

### 3.1 Characterising the excited state PESs
$C_6$-HPALD is a conformationally flexible molecule. A systemic rotor search performed in Avogadro V1.2.0 finds 7 conformational isomers which we label A-G. Ground state geometries of these conformers were then optimised with DFT/PBE0/TZVP and their analytical frequencies confirm that these geometries are local minima on the ground state PES.

Excited state properties, including energies, nuclear gradients, and nonadiabatic couplings, can be calculated accurately and efficiently with linear response time dependent density functional theory (LR-TDDFT). LR-TDDFT, like DFT, is formally exact on the condition that the true frequency-dependent exchange correlation functional is used. Its shortfalls are well documented, including its tendency to underestimate energies of states with high charge transfer character[25] or regions of the PES with strong coupling between ground and excited states.[26] Nevertheless, LR-TDDFT is widely used for nonadiabatic dynamics simulations of larger systems due to its favourable scaling with basis set size.[27] Employing LR-TDDFT for excited state dynamics, however, always benefits from a careful validation of its accuracy in comparison to high-level wavefunction methods.

To determine a method for running FSSH in the relevant region of the PES, we performed a number of excited state benchmarks at the $S_0$ geometry of the lowest energy conformer (B) of $C_6$-HPALD. A scan along the PES cross section of the -O-OH internal coordinate was initiated at the $S_0$ geometry of conformer B to validate the use of LR-TDDFT/PBE0/6-31G (calculated for 5 singlets) against MS(4)-CASPT2(10,8)/6-31G* and a number of other methods. The active space of the MS-CASPT2 calculation was selected to include the bonding and anti-bonding orbitals of the peroxide and α,β-enone chromophores, as well as the lone pairs on the oxygen atoms. All DFT/LR-TDDFT calculations in this paper were performed in Gaussian 16.[28] Ground state energies at the optimised geometries were refined with density fitted CCSD(T)-f12//cc-pVDZ-f12//def2-QZVPP in Molpro 2019.[29] MS-CASPT2 calculations were performed in OpenMolcas v18.09.[30]

The transition state ($S_1$-TS) on the $S_1$ surface was optimised using an eigenvector following Berny algorithm in Gaussian 16[28] with LR-TDDFT/PBE0/6-31G. Finding this first order saddle point on the $S_1$ surface was not a trivial task: because the seam is quite sharp, optimisation steps that were too large would cause it to fall down the steep slopes of the ridge. The $S_1$-TS geometry was verified through a vibrational frequency analysis that yielded a single imaginary frequency. An intrinsic reaction coordinate (IRC) scan was performed, initiated at this $S_1$-TS geometry. Geometries of each conformer were optimised in the $S_1$ state with the optimisation starting at their respective $S_0$ geometry. The minimum energy conical intersection (MECI) between the $S_1$ and $S_2$ states was optimised by using the search algorithm described by Harvey *et al.*.[31]

### 3.2 Predicting the photoabsorption cross section to calculate photolysis rate

The photoabsorption cross section of $C_6$-HPALD has yet to be measured experimentally. We can predict it *ab initio* by using the nuclear ensemble approach based on a harmonic Wigner distribution in the ground state[32] which captures the broadening of the spectral bands. Ground state frequencies used to generate the Wigner distribution were calculated for each conformer with DFT/PBE0/TZVP. For each of the 7 conformers, 100 nuclear configurations are sampled from their respective distribution. We calculate the absorption in the 300-400 nm range into the $S_1$ and $S_2$ electronic states separately, as well as the combined spectrum. For each sample point the vertical transitions and oscillator strengths are calculated with LR-TDDFT/PBE0/TZVP. Each peak is overlaid with a Lorentzian curve whose phenomenological broadening is set to 0.05 eV to return a continuous spectrum. The final photoabsorption cross section is a linear combination of the spectra for each conformer where the Boltzmann weights of the conformers $\omega_{conf}$ are calculated from their CCSD(T) electronic energies. We approximated the Gibbs free energy by the electronic energy because free energy corrections (from PBE0/TZVP frequencies) did not change the ordering of states.

### 3.3 Trajectory surface hopping dynamics

All trajectory dynamics simulations were performed using the following protocol, unless stated otherwise. Fewest switches surface hopping (FSSH) simulations were performed in Newton-X.[14b, 33] Energies and gradients of the first four singlet states ($S_0$-$S_3$) were calculated at each step with LR-TDDFT/PBE0/6-31G using Gaussian 09.[34] Energy based decoherence corrections were applied, as described by Granucci and Persico,[35] with the decoherence parameter α set to 0.1 a.u. Nonadiabatic coupling terms between electronic states were calculated using a time-derivative coupling scheme.[36]

The importance of the nonadiabatic effects was quantified by comparing against ab initio molecular dynamics (AIMD) in which the nonadiabatic coupling between states was set to 0, effectively restricting the trajectories to the $S_1$ state. AIMD calculations were also performed in Newton-X with identical initial conditions to the FSSH run.

Starting geometries and velocities for the trajectories were generated by randomly sampling points from a ground state Wigner distribution. For each conformer this distribution was constructed using DFT/PBE0/TZVP level normal mode frequencies at the optimised $S_0$ geometries, where a larger basis set is selected to improve the quality of the distribution. In total, we ran 250 FSSH and 50 AIMD trajectories, whose initial conditions corresponded to the Wigner distribution of conformer C. A further 109 trajectories were performed with both FSSH and AIMD, corresponding to the realistic conformer distribution where the number of trajectories corresponds to the Boltzmann weight of the conformer in the ground state. All trajectories were initiated on the $S_1$ electronic state as it corresponded to the strongest peak in the actinic region, λ > 320 nm, of the absorption cross section (available in the SI).

All trajectories were propagated up to 4 ps or until photodissociation was observed. Total energy was conserved in all trajectories up to the end point of the trajectory. Because LR-TDDFT fails to describe homolytic bond dissociation, a dissociative outcome causes a failure in energy conservation, signalling trajectory termination. Classical nuclei were propagated with a 0.5 fs time step.

### 3.4 Constructing a nonadiabatic EGME model from stationary points on the excited state PES

Each electronic structure calculation used to construct an EGME model was performed at the same level of theory (LR-TDDFT/PBE0/6-31G) as that used for surface hopping dynamics so that we might directly compare the results. Energies of all stationary points were specified with respect to the energy of the geometry optimised $S_1$ minimum which was treated as the reactant well in the model. Zero-point energy corrections were not used when defining the relative energies so as to make a direct comparison with results of dynamics calculations, in which ZPE was not rigorously constrained.

The electronic structure theory codes which we utilized provided states energies in the adiabatic ($S_0$, $S_1$, $S_2$, etc.) representation. However, as illustrated in Figure 2, the NA-EGME treats the different states in the diabatic representation (in this case an $n\pi^*$ and $n'\sigma^*$ state), and requires as input an analytical form of the diabatic states in the vicinity of the crossing point to determine $P_{ZN}$ at the seam. To derive a diabatic representation from the adiabatic energies, we considered only the coordinate along the imaginary eigenvector of the $S_1$-TS Hessian, investigating 1D motion along the 3N-7 dimensional coupling seam. The eigenvector describing this motion takes the system across the $n\pi^*/n'\sigma^*$ seam, which corresponds to extension of the peroxide bond and loss of OH, denoted by reaction coordinate R. Energies of the $S_1$ and $S_2$ adiabatic states across this coordinate are used to fit the diabatic states near the TS. We do this by constructing a simple Hamiltonian, $\mathbf{H}(R)$, which includes the two diabatic states and a coupling between them ($H_{12}$), assumed to be constant in that region.

$$\mathbf{H}(R) = \begin{pmatrix} H_{11}(R) & H_{12} \\ H_{12} & H_{22}(R) \end{pmatrix} \quad (6)$$

Analytical expressions for its two eigenvalues, $\lambda_2$ and $\lambda_2$ are determined by diagonalizing $\mathbf{H}(R)$. These eigenvalues correspond to the $S_1$ and $S_2$ adiabatic states respectively. Calculated adiabatic states were fitted to the analytical forms of the diabats given in equation 7. We assumed the dissociative state $H_{22}(R)$ to have the form of an exponential decay, and the bound state, $H_{11}(R)$, to have the form of a harmonic well.

$$\begin{aligned} H_{11}(R) &= A_{\pi*}(R - \beta_{\pi*})^2 + \varepsilon_{\pi*} \\ H_{22}(R) &= A_{\sigma*}exp(-R\beta_{\sigma*}) + \varepsilon_{\sigma*} \end{aligned} \quad (7)$$

The fitted parameters (available in the SI) were used in the NA-EGME calculation to determine $P_{ZN}$ and calculate a set of microcanonical rate constants for each energy grain.

The initial population vector $\mathbf{n}(E,t_0)$ was set up with N energy grains. To replicate the energy distribution of the wavepacket at the start of the dynamics, $\mathbf{n}(E,t_0)$ must mirror the initial conditions used in the FSSH calculations. Each initial condition sampled from the Wigner ensemble corresponds to an initial energy, a sum of its kinetic energy and its potential energy referenced to $S_1$. The distribution of total energies resembles a normal distribution, with an average initial energy above the $S_1$ minimum. This average energy corresponds to the $n_i^{th}$ energy grain in the population vector $\mathbf{n}(E,t)$ so the EGME calculations are initiated with 100% of the initial population in this grain.

We used a grain size of 50 cm$^{-1}$ in all EGME calculations and standard temperature and pressure (300K and 760 Torr) to replicate atmospheric conditions. Collision parameters used for the bath gas, He, were ε = 10.2 K and σ = 2.55 Å and the collisional energy transfer was treated using an exponential down model.[37] The collisional energy transfer parameter is set to 250 cm$^{-1}$, a value which is typical under standard atmospheric conditions.[21c] Photodissociation was assumed to be irreversible and the dissociation products were treated as a sink. To quantify the impact of nonadiabatic effects at the $n\pi^*/n'\sigma^*$ seam in these EGME calculations, the same model was re-run without allowing nonadiabatic transitions, using a standard non-equilibrium ground state EGME model. Microcanonical k(E) were calculated using RRKM theory.

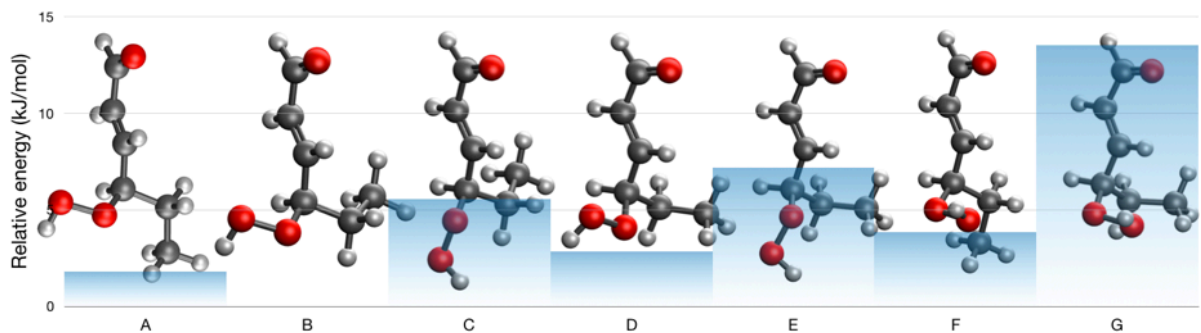

Figure 3: Geometries of the 7 rotational isomers of C$_6$-HPALD. Relative energies, shown on chart, are refined with df-CCSD(T)-f12//cc-pVDZ-f12//def2-QZVPP.

The NA-EGME model was then modified by substituting MS(4)-CASPT2/6-31G* energies at the same LR-TDDFT optimised stationary points to refine the EGME result. These calculations use the same frequencies and rotational constants calculated with LR-TDDFT/PBE0/6-31G. Single point energy calculations were performed for the following geometries: the $S_1$ minimum of the B conformer; geometries found by taking steps along the imaginary eigenvector of $S_1$-TS. For the latter case, a crossing point between the diabatic surfaces is found at a small displacement from the $S_1$-TS geometry. New diabats were fitted to the results of the scan, leading to slightly different ZN parameters.

All EGME calculations reported in this study were performed using the open source master equation solver, MESMER.[24, 38]

## 4. Results and Discussion

### 4.1 Characterising the $S_1$ and $S_2$ PESs in the Franck-Condon region and at the crossing seam

Ground state geometries and relative CCSD(T) energies of all 7 conformers are shown in Figure 3. Optimising these rotamer structures on the $S_1$ PES with LR-TDDFT converged on 7 distinct structures that maintain the orientation of the peroxide and (-CH$_2$CH$_3$) branches such that there are multiple $S_1$ minima. Conformer B remained the lowest energy conformer on the $S_1$ PES.

Our predicted photoabsorption cross section $\sigma(\lambda)$, available in the SI, indicated that the majority of the photoexcitation in the UV-Vis region is into the $S_1$ state. Integrating over $\sigma(\lambda)$, actinic flux, and quantum yield in the actinic region we can make an ab-initio estimate of the photolysis rate. Assuming a unity quantum yield we predict it to be $1.4 \times 10^{-4}$ s$^{-1}$, within a factor of three of the observed experimental rate of $6.3 \pm 0.1 \times 10^{-5}$ s$^{-1}$.[19] The cross section indicates that in the actinic region the strongest peak corresponds to absorption into the $S_1$ state.

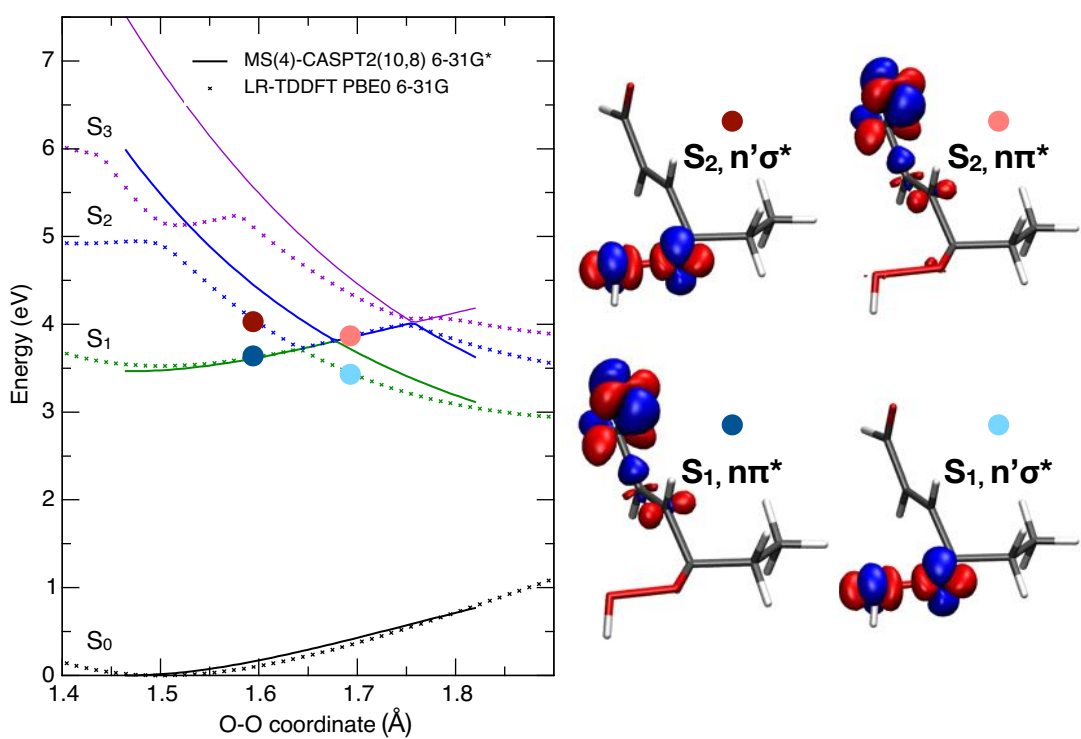

Figure 4: Energies of the first 4 excited states calculated with LR-TDDFT/PBE0/6-31G and MS(4)-CASPT2(10,8)/6-31G*, alongside electron density difference plots between $S_0$ and specified state at 2 points along the peroxide bond coordinate illustrating the change in diabatic character.

Shapes of the excited state potentials along the peroxide bond coordinate calculated with LR-TDDFT/PBE0/6-31G and MS-CASPT2(10,8)/6-31G* show good qualitative agreement. On this basis, we decided to use LR-TDDFT/PBE0/6-31G as the electronic structure method for all FSSH and AIMD calculations in this paper. Rigid scans along the peroxide bond dissociation coordinate in Figure 4 show a near degenerate region between the $S_1$ and $S_2$ states at 1.65 Å, and between $S_2$ and $S_3$ states at 1.76 Å. Benchmarks using larger basis sets show that as the bond extends beyond 1.75 Å, LR-TDDFT provides a poor description of homolytic dissociation. This, however, will not be a significant problem for the dynamics because the $S_1$ potential is dissociative beyond this point, at which point trajectories were terminated.

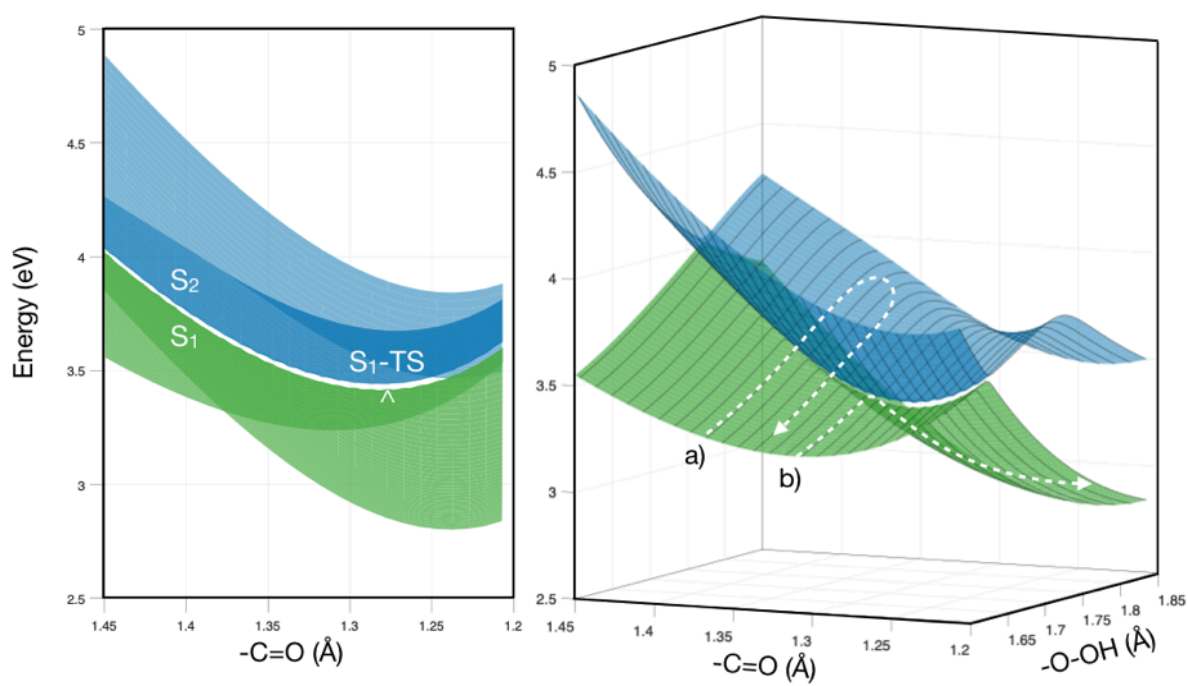

Figure 5: Scan of the $S_1/S_2$ nonadiabatic seam initiated at $S_1$-TS, excitation energies calculated with LR-TDDFT/PBE0/6-31G. Left panel highlights the extended near-degenerate (3N-7) seam. a) Diabatic trapping mechanism; b) Adiabatic passage across seam leading to loss of OH.

In the density difference plots shown in figure 4, it's visible that the $n\pi^*$ transition which characterises the $S_1$ state in the bound region of the PES (O-O extended to 1.6 Å) is located mostly on the α,β-enone chromophore. At the same geometry, the $n'\sigma^*$ transition to the $S_2$ state is located mostly along the -O-OH bond. When the peroxide bond is extended to 1.7 Å the ordering of the diabatic states swaps, such that the $S_1$ state is characterised by the $n'\sigma^*$ transition. The region of strong nonadiabatic coupling observed between the $S_1$ and $S_2$ at 1.65 Å in Figure 4 is a single point on an extended (3N-7) seam where the $n\pi^*$ and $n'\sigma^*$ diabatic states cross. We located critical points along this seam which included an $S_1/S_2$ MECI as well as a saddle point on the $S_1$ surface ($S_1$-TS) which is the minimum energy geometry in the space of this seam. The energy of the MECI is 31.3 kJ mol$^{-1}$ above the $S_1$ minimum of the lowest energy conformer, B. The energy of the $S_1$-TS is at 16.1 kJ mol$^{-1}$ relative to conformer B. Visualising the normal mode corresponding to the only imaginary frequency at the $S_1$-TS showed synchronization between stretching in the -O-OH coordinate and compression in the C=O coordinate. This highlights that this mode couples the α,β-enone and peroxide chromophores and therefore will be important for describing the reaction coordinate. Furthermore, the remarkably high value of the imaginary frequency ($v_{im}$ = -3534.1 cm$^{-1}$) illustrates the sharpness of the (3N-7) seam in the vicinity of the $S_1$-TS. To visualise this seam we performed a rigid 2D scan along the -C=O and -O-OH stretching coordinates of $C_6$-HPALD that correspond to the two coupling chromophores. Results of this scan are shown in fFigure 5.

An intrinsic reaction coordinate (IRC) scan initiated at the $S_1$-TS geometry converges on the dissociated structure and the $S_1$ minimum of conformer C, as can be seen in Figure 6. Energies, frequencies, and rotational constants at these critical points are tabulated in the SI.

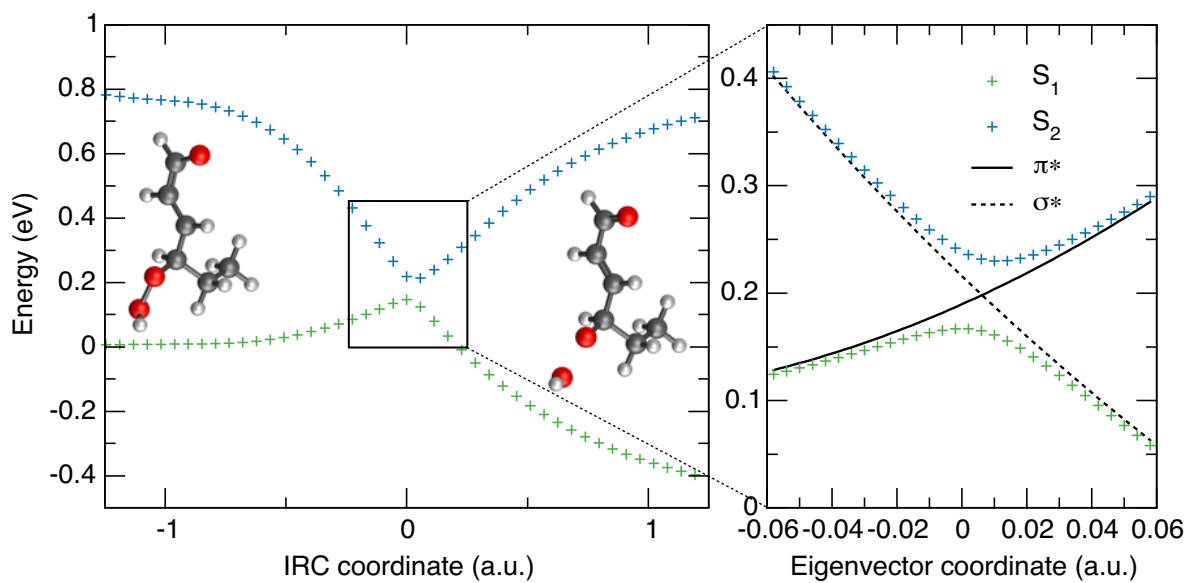

Figure 6: Intrinsic reaction coordinate scan started at the $S_1$-TS. Energy is shown relative to the $S_1$ minimum of conformer C. Geometries are shown at the terminal step of the IRC scan. Right panel shows a scan across the TS geometry along the eigenvector of the imaginary normal mode. Fitted diabatic states used in the NA-EGME calculation are shown in black.

## 4.2 Direct comparison between NA-EGME and FSSH for a single conformer

Our exploration of the excited state PES located a direct reaction coordinate between the $S_1$ minimum of conformer C and the $S_1$-TS, shown by the IRC in Figure 6. In the following section, we consider a simple photodissociation model based on a potential well ($S_1$-C) and a single barrier ($S_1$-TS) linked by this reaction coordinate, that ignores all other $C_6$-HPALD conformations.

### 4.2.1 Description of the seam crossing and OH loss rate from FSSH dynamics

We began by running 50 AIMD and FSSH trajectories, whose initial conditions were sampled from the ground state Wigner distribution of conformer C. The two sets of trajectories shown in Figure 7 are projected on to the -C=O and -O-OH coordinates, illustrating the passage of trajectories across the seam. By observing HPALD dynamics prior to dissociation we see that the molecule remains in the $S_1$ potential well for a number of vibrational periods and explores the available phase space within its initial conformation.

For all 50 FSSH trajectories the net adiabatic population remained largely on the $S_1$ state, with 90% hopping to the $S_2$ state at some point during the run. Only a single trajectory hopped to the $S_3$ state, and no population on $S_0$ was ever observed. On this timescale we expect that the dynamics are limited to the $S_1$ and $S_2$ adiabatic states. In the Franck-Condon region of the PES the $S_1$ state exhibits $n\pi^*$ character and is near an $S_1$ PES minimum. AIMD trajectories in Figure 7 indicate that for loss of OH to occur, the -O-OH coordinate must extend in concert with the compression of the -C=O bond, causing the $S_1$ state to change character from predominantly $n\pi^*$ to the dissociative $n'\sigma^*$ character. The wavepacket must necessarily proceed across the seam adiabatically, however hopping to the non-dissociative $S_2$ state can occur.

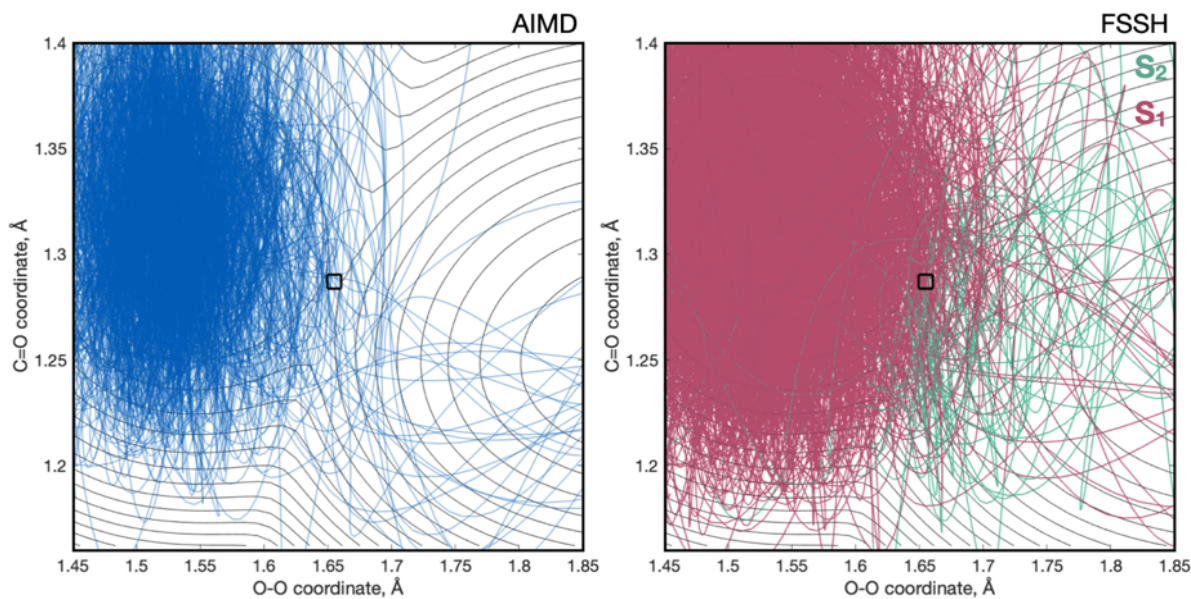

Figure 7: Projection of 50 AIMD and FSSH trajectories of conformer C into the -C=O and -O-OH coordinates. FSSH panel (right) illustrates the switch from $S_1$ to $S_2$ with a colour change. Background contour plot shows the shape of the $S_1$ PES from a rigid scan initiated at the $S_1$-TS (indicated by the black square) then scanned across these two coordinates, illustrating the $S_1$ potential energy well where the trajectories are initialised, and the dissociative potential on the other side of the barrier.

In Figure 7, we observe that while many FSSH trajectories that travel across the barrier rebound back towards the $S_1$ well, almost all AIMD trajectories which cross the barrier dissociate. In the nonadiabatic case such motion visibly corresponds with a switch to the $S_2$ state. This mechanism is referred to as diabatic trapping (originally described by Martínez *et al.* as up-funnelling) whereby a trajectory remains on the same diabatic state as it crosses the coupling region thereby preserving its electronic character.[39] Because of this, OH loss is faster for AIMD trajectories since crossing this (3N-7) seam will necessarily lead to a dissociative outcome, whereas in FSSH the trajectory might become trapped in $S_2$ and rebound instead. A similar upwards hopping process is observed in the work of Blancafort *et al.* in the bis-adamantyl radical cation that contains two weakly coupled chromophores.[40] We note some similarities between their system and ours, such as the extended near-degenerate seam between two adiabatic states. Qualitatively, we note that diabatic trapping is likely in systems where the CI branching space vectors are of significantly different magnitude, as is the case here (CI branching space vectors available in the SI).

Next, we consider the rate of dissociation as determined by the dynamics. Three possible outcomes have been observed in the FSSH results: loss of OH (38 trajectories), loss of $HO_2$ (9 trajectories), and no dissociative reaction (3 trajectories). The corresponding AIMD results are as follows: loss of OH (45 trajectories), loss of $HO_2$ (3 trajectories), and no dissociative reaction (2 trajectories). A dissociative outcome is defined as the extension of either the C-OOH or O-OH bond coordinate beyond 1.75 Å and 1.9 Å respectively. Benchmark scans of the PES along these coordinates have shown a potential barrier at 1.65 Å, beyond which the molecule is unlikely to recombine. Dissociating trajectories terminated soon after this nonadiabatic barrier is crossed due to the unreliability of LR-TDDFT in its description of homolytic dissociation. These trajectories are included in the analysis up to the point of dissociation since we can assume that once the bond has extended beyond the threshold, the rate coefficient for reassociation is very small. Loss of $HO_2$ is a minor dissociative channel which has been suggested experimentally for other peroxides.[41] Its mechanism in $C_6$-HPALD appears to be linked with diabatic trapping because all FSSH trajectories terminating in this way show an $S_2$ to $S_1$ hop 20 fs prior to dissociation. Given its low probability, and because it cannot be treated with a kinetic model, the $HO_2$ loss channel is excluded in the following analyses.

To ensure that the FSSH result is converged we ran another 200 trajectories by using the same 50 initial conditions but inserting a new random seed for the surface hopping algorithm 4 times. In Figure 8 we see that the results are well converged with as few as 50 trajectories. Biexponential fits of HPALD population decay are available in the SI. A biexponential least-squares fit indicates that there are two separate decay timescales.[42] The fast decay corresponds to trajectories that dissociated ballistically (OH loss takes less than 200 fs), while others remained in the pre-dissociative $S_1$ well until the trajectory was able to cross the seam allowing more time for intra-vibrational relaxation to occur. Decay constants for the slow fraction of the decay are 1.87 ps for FSSH and 1.29 ps for AIMD.

**4.2.2 Calculating rate of OH loss using an NA-EGME model**
Our NA-EGME model assumes that to describe the primary photodissociation channel leading to the loss of OH we need only to consider the $n\pi^*/n'\sigma^*$ state coupling along a 1-D coordinate over the top of the TS. This necessitated only a normal mode analysis at the $S_1$ minimum of conformer C ($S_1$-C) and the $S_1$-TS barrier which is energetically 14.13 kJ mol$^{-1}$ higher. The optimised MECI is 15.2 kJ mol$^{-1}$ above the $S_1$-TS. The 3N-7 dimensional geometry of the nonadiabatic seam means the MECI is unlikely to be an important critical point since the wavepacket does not need to pass through it to reach the $S_2$ state. For this reason, the MECI's influence on the dissociation rate can be neglected and we choose not to include it in the model. Total energies of the 50 initial conditions are normally distributed with the average initial energy at 453.4 kJ mol$^{-1}$ above the $S_1$-C minimum. This energy corresponds to the 758th grain in the population vector $\mathbf{n}(E,t)$ and so, the EGME calculations were initiated with 100% of the population in this energy grain.

The HPALD decay rates calculated using A-EGME and NA-EGME are presented in Figure 8, illustrating that the photolysis rates obtained with the trajectory-based approaches are quantitatively similar to those obtained from EGME models. Including nonadiabatic effects slows down the decay rate approximately 6-fold ($\tau_{NA-EGME}$ = 2.72 ps) in comparison to the rate calculated when nonadiabaticity is neglected ($\tau_{A-EGME}$ = 0.45 ps). The decay rate is shown to be robust to the initial energy grain distribution and small variations in frequencies by the sensitivity analyses provided in the SI.

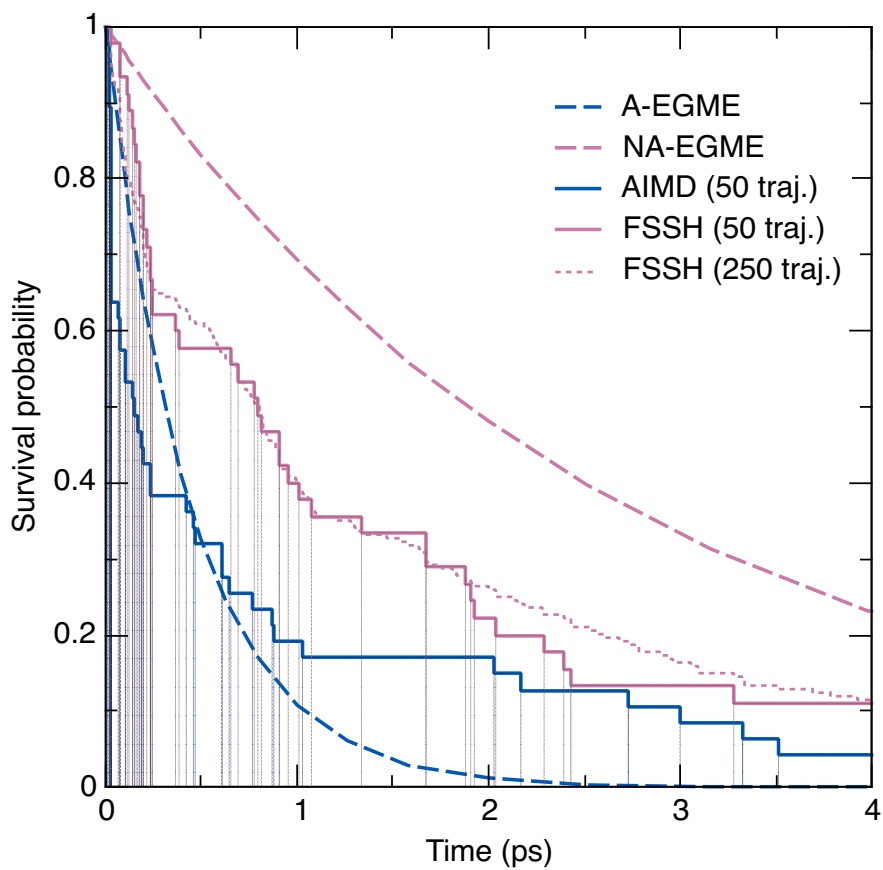

Figure 8: Comparison of $C_6$-HPALD decay rate between EGME and trajectory-based methods for conformer C. Convergence of FSSH at 50 trajectories indicated by the similarity to the 250-trajectory result.

### 4.3 Comparison of the two extended models, including all conformers

Having established that treating the photodissociation of conformer C in isolation shows good agreement between NA-EGME and FSSH results we can extend this simple model to include all conformers of $C_6$-HPALD shown in Figure 3. For the dynamics calculations we simply projected the ground state conformer distribution into the excited state such that the set of trajectory initial conditions was representative of the rotamer distribution. Boltzmann weights of the A-G conformers (listed in the SI) determine the number of trajectories to be run for each. For TST-type methods like EGME, molecular torsions can be challenging because the rigid rotor approximation breaks down due to the highly anharmonic hindered rotor modes. Ideally, each conformation and its corresponding TS should be treated separately.[43] However, for 7 rotational conformers this approach would necessitate a cumbersome search for 30 separate TSs in a 3N dimensional phase space. Instead, we propose a pared down model that uses the global conformer minimum ($S_1$-B) and the $S_1$-TS to calculate the OH loss rate.

#### 4.3.1 Conformational changes and realistic dynamics of OH loss in FSSH trajectories

The relative numbers of trajectories initiated at each conformer corresponds to their Boltzmann weight in the ground state calculated using CCSD(T) energies: A: 24; B: 50; C: 5; D: 16; E: 3; and F: 11.

Dihedral angles $\phi_1$ and $\phi_2$ can be used as a shorthand to distinguish the conformers over the course of a trajectory. This can be seen in Figure 9 which shows all of the dissociating trajectories exploring the rotational phase space and highlighting that the timescale of conformer interconversion is comparable to that of OH loss. Trajectories corresponding to conformers C and F especially tend to remain conformationally locked, supporting our previous assumption that conformer C could be treated independently. There is a flux of trajectories from conformer B to C suggesting that the rotational barrier between them is small.

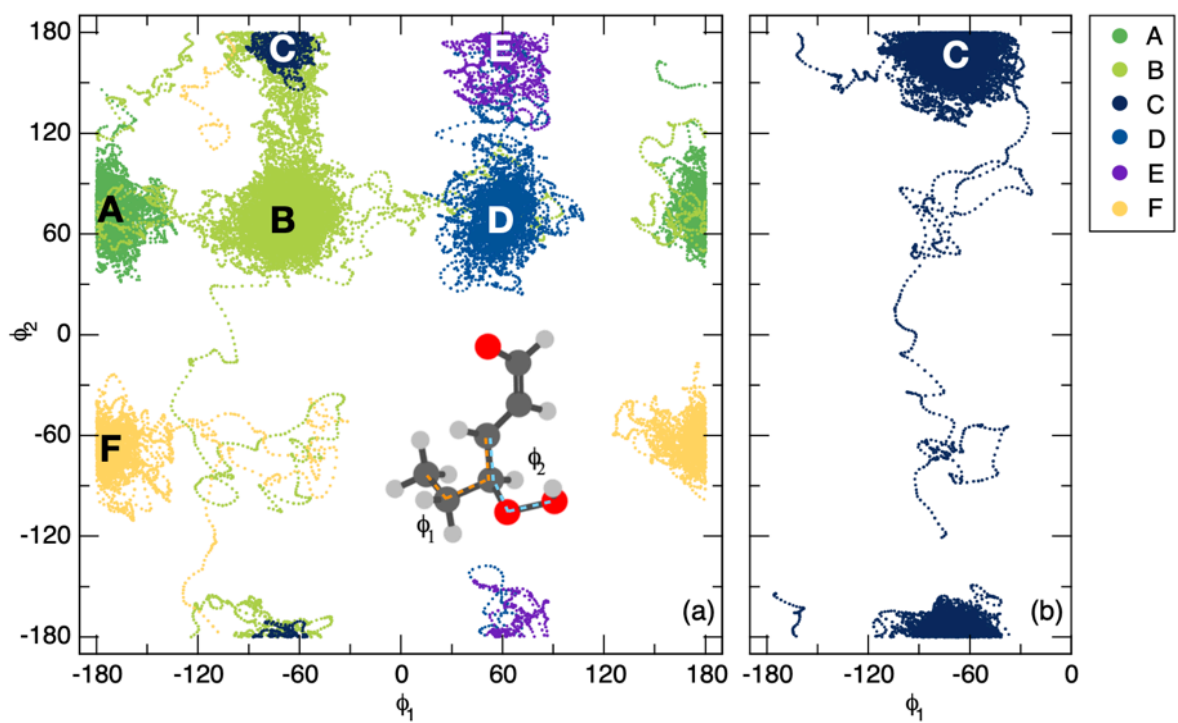

Figure 9: HPALD conformer interconversion over a 4 ps timescale. Left panel shows the evolution of the two dihedral angle coordinates $\phi_1$ and $\phi_2$ that define the conformation of the HPALD molecule at a given time step for all 109 FSSH trajectories. Right panel shows 50 trajectories of conformer C only.

Of the 109 FSSH trajectories initiated on the $S_1$ state we observed the following outcomes: loss of OH (50 trajectories), loss of $HO_2$ (15); no dissociation (43). The corresponding results of the AIMD simulations are: loss of OH (82), loss of $HO_2$ (10), no dissociation (17). The 95% margin of error shown by the error bars in Figure 10c illustrates that the difference between dissociative outcomes in adiabatic vs. nonadiabatic simulations is significant.

Analysis of the mean adiabatic population shows that on average population remained on the $S_1$ surface, rarely falling below 95%, with a fraction of population moving into the $S_2$ state. Very few trajectories hopped into the $S_3$ state and no population of the $S_0$ state was observed on the timescale of the simulation. The survival probability is fitted to an exponential decay, with a first order lifetime, $\tau_{FSSH}$, of 4.6 ps. AIMD dynamics of OH loss are better fitted to a double exponential, shown in Figure 10d. Approximately a quarter of trajectories are dissociated on a fast timescale with a lifetime, $\tau_{fast}$, of 58 fs. The rest of the trajectories dissociate on a similar timescale to the FSSH simulations with lifetime, $\tau_{slow}$, of 2.4 ps.

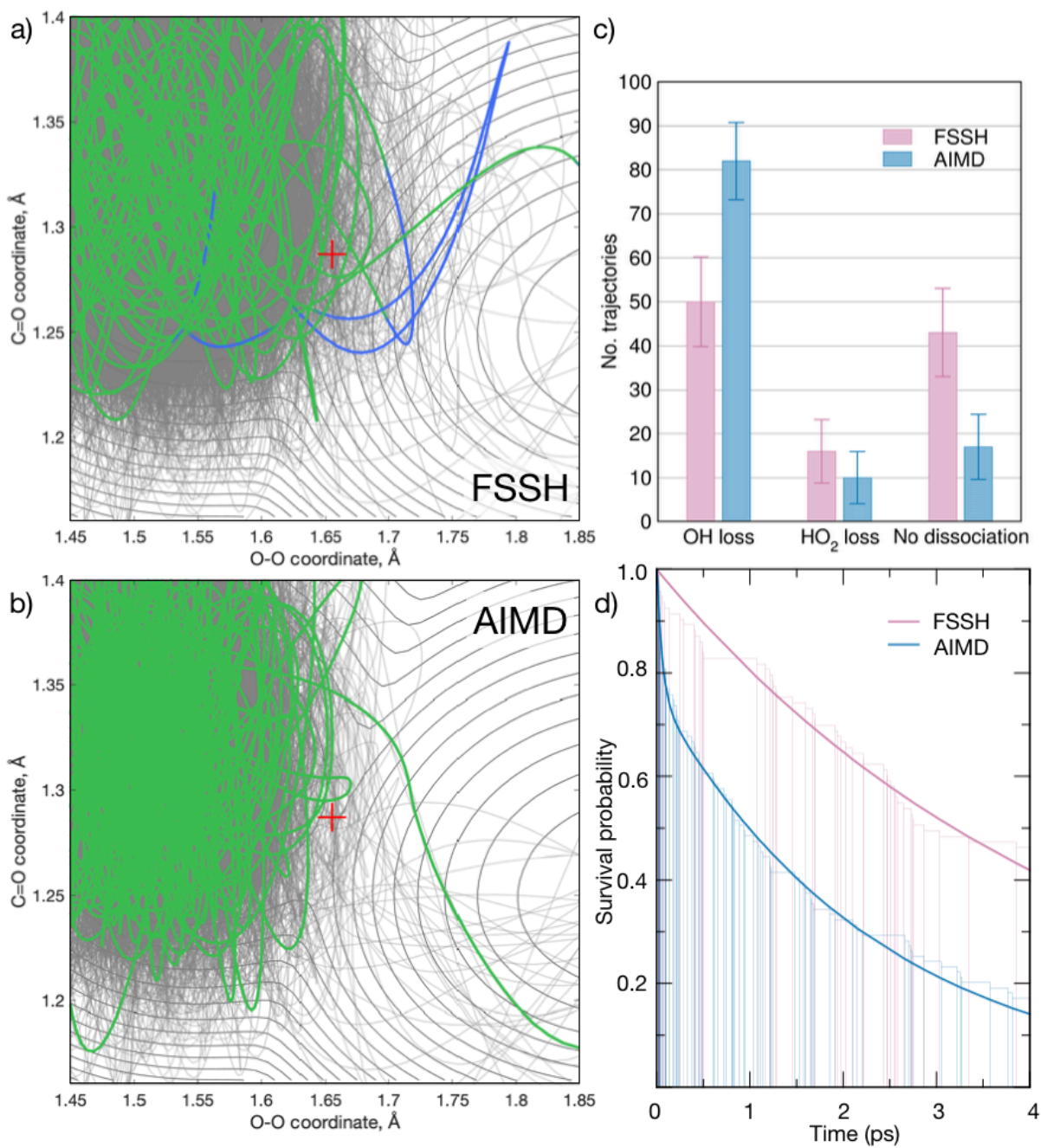

Figure 10: Dynamics results for all conformers of $C_6$-HPALD. a) An example of a diabatically trapped trajectory where the switch to $S_2$ is illustrated by a colour change. b) Example AIMD trajectory that moves in the $S_1$ well for a number of vibrational periods before crossing the barrier and dissociating immediately. c) Outcomes of the 109 trajectories with 95% confidence intervals showing that differences between FSSH and AIMD are significant. d) Survival probabilities with respect to OH loss, fitted to exponential decay functions.

#### 4.3.2 An approximate all-conformer NA-EGME model for OH loss

Now that we have the results of trajectory dynamics initiated from a realistic ground state conformer distribution we construct a new, more realistic, NA-EGME model. Since a large fraction (50 out of 109) of the trajectories were initiated from the conformer B initial condition we use its $S_1$ minimum as the reactant well ($S_1$-B) and the $S_1$-TS. The average initial energy of these trajectories was at 487.0 kJ mol$^{-1}$ above the $S_1$-B minimum corresponding to the 814th energy grain in the population vector

**n**(E,t). Results of the NA-EGME calculations based on this model are shown in Figure 11. We see that including nonadiabaticity once again has a strong effect on the microcanonical rate coefficients shown in the inset. The nonadiabatic lifetime $\tau_{ZN}$ is 1.7 ps, once again ~6x greater than the adiabatic lifetime, $\tau_{RRKM}$ = 0.3 ps.

We have tested the importance of torsional anharmonicity using the hindered rotor approach, as implemented in MESMER. This sensitivity analysis ensures that the presence of anharmonic rotational modes does not significantly alter the ratio of densities of states, and the corresponding decay constants. Normal mode frequencies corresponding to torsional motion were projected out of the Hessian.[44] Results of rigid torsional scans performed over 4 torsional bonds at the $S_1$-B minimum and $S_1$-TS geometries were input into the MESMER calculation. Results available in the SI show that incorporating these torsional effects does not significantly impact the reaction profiles. We suggest this could be due to the similarity between torsional profiles at the $S_1$ minimum and TS geometries that result in a cancellation of errors.

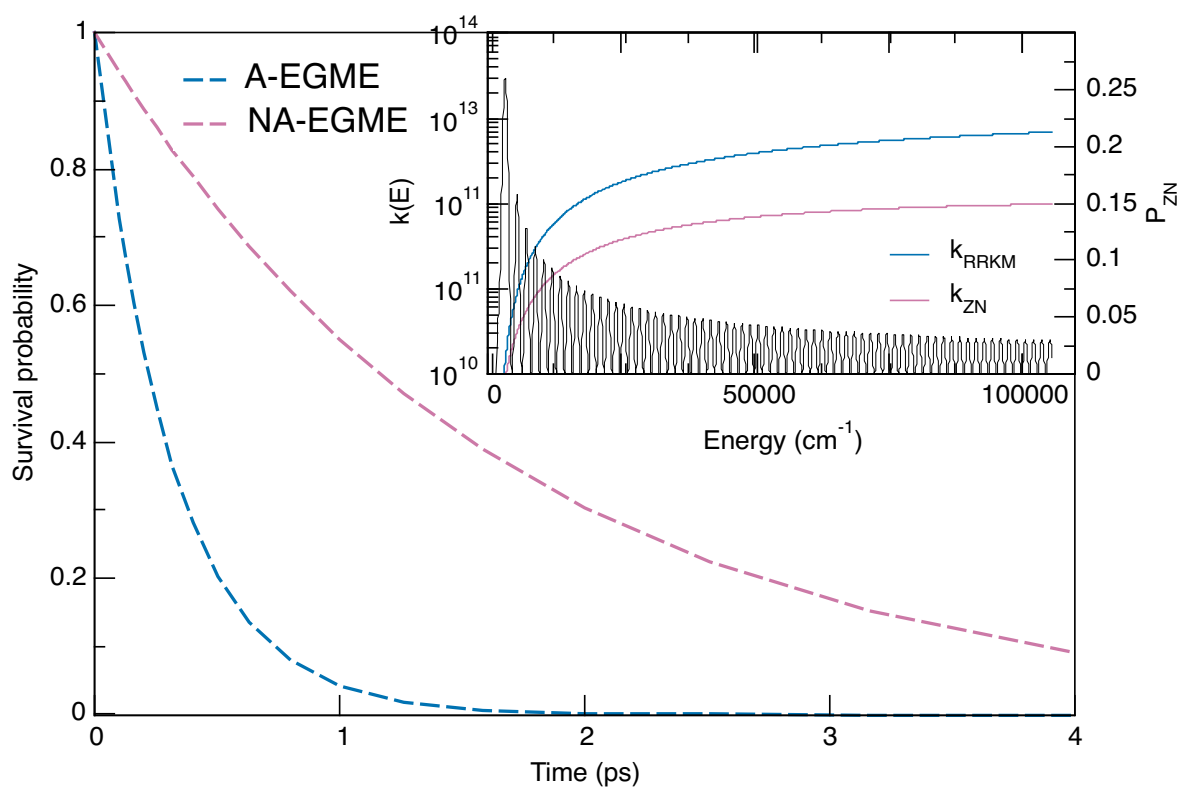

Figure 11: Results of EGME calculations, with microcanonical rates calculated using an adiabatic (A-EGME) and a nonadiabatic (NA-EGME) expression. Inset shows the energy resolved microcanonical rate constants ($k_{RRKM}$ and $k_{ZN}$) and the ZN transition probability (black line).

An implicit assumption in this treatment of hindered rotations is that rotamer interconversion is fast on the reaction timescale. However, conformational analysis of the trajectories in Figure 9 shows that while internal rotations are present, they are not fast. For this simplified model we make an approximation to consider only the global $S_1$ minimum ($S_1$-B) and a single lowest point on the (3N-7) seam ($S_1$-TS). Similarity between the frequencies and rotational constants of the conformers suggest that this is an acceptable compromise in this case.

### 4.3.3 Correcting the NA-EGME model with MS-CASPT2 energies
Our NA-EGME model could be improved by the addition of all inter-conformer transition states. However, the need to search for each critical point on a 3N dimensional PES can undermine the simplicity of the approach proposed here. Instead, we can exploit the comparatively low computational cost of the EGME calculations by using energies calculated with a more sophisticated, multireference, electronic structure method at the stationary points. By assuming that the locations of the stationary points optimised with LR-TDDFT give a broadly accurate representation of the PES, the model can be adjusted by using MS(4)-CASPT2(10,8)/6-31G* energies (with LR-TDDFT frequencies and rotational constants) to calculate the population profile of HPALD over time. Results of this scan are shown in Figure *12*. New parameters are used in the ZN equations, based on the fitting of diabats to new MS-CASPT2 energies calculated across the same eigenvector coordinate as in the earlier calculation.

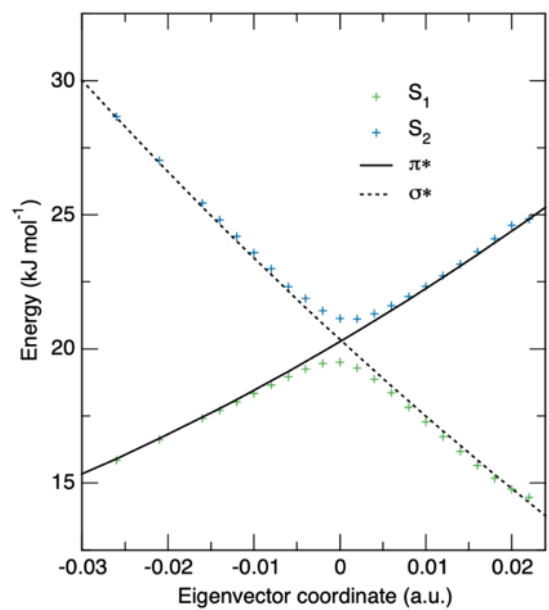

Figure 12: MS(4)-CASPT2(10,8)/6-31G* scan across the $S_1$-TS geometry along the eigenvector of the imaginary normal mode, optimised with LR-TDDFT/PBE0/6-31G. Fitted diabatic states used for parameter fitting in the NA-EGME calculation are shown in black

Results of NA-EGME calculations based on MS-CASPT2 energies show that the difference between the nonadiabatic and adiabatic rates is more significant than that produced using TDDFT/PBE0 energies. The adiabatic lifetime of HPALD is $\tau_{A\text{-EGME}} = 0.44$ ps whereas the nonadiabatic lifetime, $\tau_{NA\text{-EGME}} = 30$ ps, is 70 times slower. Nonadiabatic coupling between the states is weaker increasing the likelihood of transition to the $S_2$ state, as can be seen in the SI.

## 5 Discussion

The most direct comparison between the two approaches can be seen in the models that isolate conformer C. Figure *8* demonstrates the remarkable, almost quantitative, similarity between the results of trajectory-based methods and EGME: the FSSH photolysis lifetime $\tau_{FSSH} = 1.9$ ps is comparable to the NA-EGME lifetime $\tau_{NA\text{-EGME}} = 2.7$ ps. The impact of nonadiabatic effects on the photolysis rate is stronger in the EGME results: the calculated adiabatic lifetimes are $\tau_{A\text{-EGME}} = 0.5$ ps and $\tau_{A\text{-EGME}} = 1.29$ ps. A direct comparison between the decay rates in the extended models that include all conformers can be seen in Figure 13. Nonadiabatic EGME and FSSH methods returned lifetimes differing by less than factor of 3: $\tau_{NA\text{-EGME}} = 1.7$ ps and $\tau_{FSSH} = 4.6$ ps. These results are still qualitatively similar, which is remarkable given the stark differences between the two approaches. For the adiabatic simulations the comparison is made with the slow component of the fitted decay, with $\tau_{A\text{-EGME}} = 0.31$ ps and $\tau_{AIMD} = 2.4$ ps, out by less than a factor of 8. Trajectory surface hopping simulations indicate that a diabatic trapping mechanism is responsible for this deceleration as it causes the nuclear wavepacket to be trapped in a bound diabatic state, preventing direct dissociation. To quantify the impact of nonadiabatic effects at the seam, for the both the dynamical and the master equation approaches, we compare the ratios of the nonadiabatic/adiabatic lifetimes. This value is 1.9 for the trajectory methods, and 5.5 for the kinetic model. The difference may arise in part from the assumption within the EGME model that allows for only a single seam crossing. It is also important to highlight that the excited-state dynamics performed here assume the formation of a nuclear wavepacket upon light absorption. Such initial condition corresponds to a scenario where the molecule is photoexcited by an ultrashort laser pulse, rather than continuous irradiation with sunlight as it would happen under atmospheric conditions. The question of selecting proper initial conditions for the excited-state dynamics of atmospheric molecules is discussed further in Suchan *et al.*.[45] A protocol aiming at simulating sunlight absorption processes was also recently proposed.[46] In the context of this work, this assumption does not affect the comparison between the EGME models and the dynamics, but does limit the claims we can make about the atmospheric implications of our results.

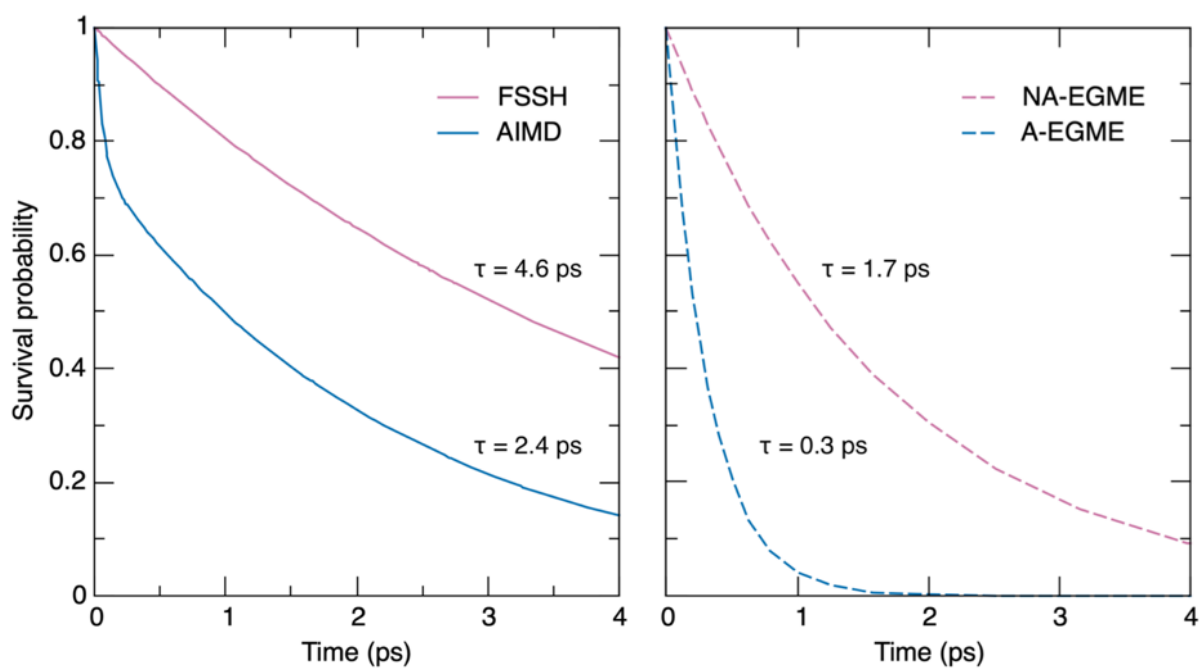

Figure 13: Side-by-side comparison of the two models used to calculate the dissociation rate for $C_6$-HPALD. The fit to the population decay is presented for both FSSH and AIMS.

In this paper we aimed to validate a nonadiabatic EGME model against FSSH by calculating the rate of OH loss in $C_6$-HPALD, which has been experimentally investigated by Wolfe *et al.*.[19] We highlight the approximations made in the construction of the master equation model and outline how this model can be improved. Many features of the $C_6$-HPALD dissociation process seem to justify these approximations. This includes the picosecond timescale of photodissociation; ease of energy exchange between the many modes of HPALD; the (3N-7) geometry of the seam which makes the 1-D seam crossing model appropriate. Of course, the exploratory value of running dynamics simulations cannot be superseded by a model that requires existing knowledge of important stationary points. Without performing the FSSH calculations the diabatic trapping mechanism would not have been identified. Nonadiabatic transitions are ultimately caused by nuclear motion and so atomistic simulations are necessary for an accurate description of wavepacket dynamics. However, when nonadiabatic transitions between excited states occur on a slow timescale we are limited by the computational cost of running long trajectories and using the NA-EGME model allows us to refine the energies of the critical points whilst reproducing the overall impact of diabatic trapping on the photolysis rate. A shorthand calculation of CPU core hours needed to execute both types of calculations tells us that running dynamics cost 200 times more than the NA-EGME approach, including all the preliminary electronic structure calculation of the parameters needed for the model. In instances of slow photodynamic reactions with known mechanisms, alternative models might be explored before choosing to run trajectory dynamics.

The workflow to perform a nonadiabatic EGME analysis on this type of crossing can then be summarised as follows.
1. Locate and characterise the critical points on the excited state PES. These include the bound minima near the FC region, the conical intersections, and the adjacent transition states.
2. Identify the normal mode at the crossing point that corresponds to the exciton moving from one chromophore to the other.
3. Perform a scan across this normal mode and fit diabatic curves to the shape of the crossing point along a 1-dimensional analytical model.
4. Construct an EGME model of the seam crossing, using the fitting parameters obtained in step 3 for ZN transition probabilities. This utility is currently implemented in MESMER.

These findings describe a type of crossing between adiabatic surfaces that is intermediate to the traditional representation of a two-cone type conical intersection (Figure 14, left panel) and a fully degenerate seam one might see in the context of intersystem crossing (Figure 14, right panel). The protocol described here is applicable when calculating rates for this type of trivially unavoided crossing i.e. when collapsing the reaction coordinate to a single dimension is appropriate. These coordinates could be identified through principal component analysis of trajectory dynamics.

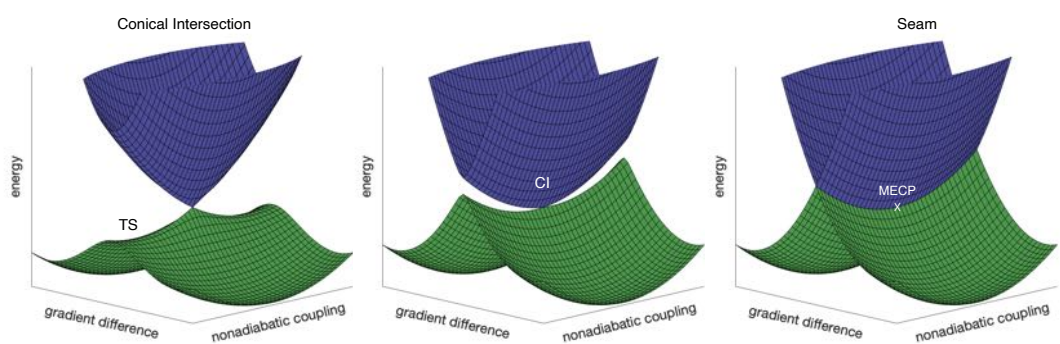

Figure 14: Types of crossings between adiabatic states, the geometry of the intersection is determined by the extent of the nonadiabatic coupling between states. Central panel represents a model topology like that of the $n\pi^*/n'\sigma^*$ crossing in $C_6$-HPALD where the CI is only a single point on an extended seam and so becomes less important in the overall description of the nonadiabatic transition.

# 6. Conclusions

We directly compared the performance of FSSH dynamics to that of a nonadiabatic EGME model by conducting two side-by-side studies of $C_6$-HPALD photodissociation. Both methods establish that the nonadiabatic coupling at the extended seam is significant and reduces the rate of OH loss. The lifetimes of $C_6$-HPALD based on these fundamentally different models indicate that a reduced dimensionality NA-EGME treatment for avoided crossings can reproduce results of dynamics to within an order of magnitude. Further work is needed to investigate the rate of intra-vibrational relaxation between all modes, so as to determine the exact limits of the regime where this kind of protocol can be applied. Similarity between the dynamic and EGME results also raises the question of timescale, since intra-vibrational relaxation must be fast to satisfy the key assumption of RRKM theory. It is unclear whether this is satisfied in this case, and so merits further work to investigate the energy redistribution between modes prior to the dissociation. Some purely dynamical features such as loss of $HO_2$ could not be included in an EGME treatment, and merit further exploration to determine the significance of $HO_2$ loss to the atmospheric mechanism. Alongside the significant improvement in computational cost we highlight that approaching this photolysis mechanism from both the kinetic and the dynamic perspective offers insights into different aspects of the dissociative process.

# Acknowledgements

D.R.G. acknowledges funding from the Royal Society as a University Research Fellow. RJS is supported by EPSRC Programme Grant No. EP/P021123/1. D.S. acknowledges PhD studentship support from the EPSRC Centre for Doctoral Training in Theory and Modelling in Chemical Sciences (EP/L015722/1). This project has received funding from the European Research Council (ERC) under the European Union's Horizon 2020 research and innovation programme (Grant Agreements 803718 and 701355 – Projects SINDAM and NAMDIA). Thanks to Prof. Todd Martinez for useful discussions at various stages.

# Supporting Information for

# Dynamics in multi-chromophore systems: Nonadiabatic versus energy grained master equation treatments

Darya Shchepanovska, Robin J. Shannon, Basile F. E. Curchod*, and David R. Glowacki*
*glowacki@bristol.ac.uk; basile.f.curchod@durham.ac.uk

## S1. Optimised geometries of C$_6$-HPALD conformers

The chemical structure of the C$_5$-HPALD proxy, C$_6$-HPALD, was taken from Wolfe et al.[1] Geometry optimisations of the (E)-4-Hydroperoxyhex-2-enal structures were initialised from the 7 distinct conformers A-G (2 rotatable bonds) generated by the systematic rotor search algorithm in Avogadro V1.2.0[2]. The resulting conformers were pre-optimised in Avogadro with the UFF forcefield via the steepest descent algorithm.

The geometry optimizations for all 7 conformers were then performed with LR-TDDFT (PBE0/TZVP) as implemented in Gaussian 16[3] with frequencies calculated at the same level.

Optimised geometries of all 7 conformers are presented here (in Angstrom):

A

| | | | |
|---|---|---|---|
| C | 2.9961760000 | -1.2969820000 | 0.1292600000 |
| C | 1.5313620000 | -1.2598110000 | -0.2770690000 |
| H | 3.5474400000 | -0.4583130000 | -0.2981830000 |
| H | 3.4662400000 | -2.2203780000 | -0.2156750000 |
| H | 3.1048560000 | -1.2548880000 | 1.2167460000 |
| H | 1.4330740000 | -1.3172610000 | -1.3655990000 |
| C | 0.7986030000 | -0.0108520000 | 0.2065270000 |
| H | 1.0004470000 | -2.1241120000 | 0.1316410000 |
| O | 1.4325020000 | 1.0732750000 | -0.4568590000 |
| H | 0.9314020000 | 0.1030380000 | 1.2907210000 |
| C | -0.6546380000 | -0.0702540000 | -0.1306310000 |
| C | -1.6324830000 | -0.2126420000 | 0.7623700000 |
| C | -3.0474550000 | -0.3124890000 | 0.3605390000 |
| O | -3.4399140000 | -0.3389510000 | -0.7808060000 |
| H | -0.9108440000 | -0.0116850000 | -1.1870720000 |
| H | -1.4189440000 | -0.2445090000 | 1.8280760000 |
| H | -3.7718670000 | -0.3625550000 | 1.1993780000 |
| O | 0.8811420000 | 2.2867190000 | 0.0715100000 |
| H | 1.6789750000 | 2.7004970000 | 0.4232240000 |

B

| | | | |
|---|---|---|---|
| C | -1.2915260000 | 2.4517500000 | 0.2002250000 |
| C | -1.7351460000 | 1.1407600000 | -0.4277020000 |
| H | -1.4939960000 | 2.4670250000 | 1.2747310000 |
| H | -1.8278840000 | 3.2895140000 | -0.2494580000 |
| H | -0.2227680000 | 2.6255210000 | 0.0589530000 |
| H | -2.8118440000 | 1.0018510000 | -0.2930580000 |
| C | -1.0430520000 | -0.0789720000 | 0.1781110000 |
| H | -1.5487480000 | 1.1488550000 | -1.5055890000 |
| O | -1.7228770000 | -1.1889690000 | -0.3914370000 |
| H | -1.1851750000 | -0.0854660000 | 1.2670720000 |
| C | 0.4116430000 | -0.1159770000 | -0.1514400000 |
| C | 1.3956710000 | 0.0599570000 | 0.7287610000 |
| C | 2.8136420000 | 0.0521150000 | 0.3245380000 |
| O | 3.2053030000 | -0.0326240000 | -0.8142220000 |
| H | 0.6649910000 | -0.2818700000 | -1.1973500000 |
| H | 1.1846850000 | 0.1977140000 | 1.7863600000 |
| H | 3.5408680000 | 0.1297690000 | 1.1588080000 |
| O | -1.1855550000 | -2.3790760000 | 0.2030320000 |
| H | -1.9824880000 | -2.7453530000 | 0.6055990000 |

C

| | | | |
|---|---|---|---|
| C | -0.9626170000 | 2.4087570000 | 0.0493740000 |
| C | -1.4524880000 | 1.0798020000 | -0.5019210000 |
| H | -1.1915480000 | 2.5023280000 | 1.1148550000 |
| H | -1.4456830000 | 3.2416910000 | -0.4652870000 |
| H | 0.1168130000 | 2.5243320000 | -0.0719710000 |
| H | -2.5378360000 | 1.0124430000 | -0.3900300000 |
| C | -0.8530890000 | -0.1286900000 | 0.2153540000 |
| H | -1.2292260000 | 0.9996760000 | -1.5710370000 |
| O | -1.4467810000 | -1.3541840000 | -0.2104690000 |
| H | -0.9972300000 | -0.0278350000 | 1.2981760000 |
| C | 0.5943750000 | -0.2953170000 | -0.1005200000 |
| C | 1.5909510000 | -0.0460360000 | 0.7472720000 |
| C | 3.0034310000 | -0.1856370000 | 0.3427830000 |
| O | 3.3755010000 | -0.4512050000 | -0.7742270000 |
| H | 0.8387060000 | -0.6152530000 | -1.1120480000 |
| H | 1.3971450000 | 0.2544590000 | 1.7735930000 |
| H | 3.7428510000 | -0.0311660000 | 1.1548300000 |
| O | -2.7959590000 | -1.3651860000 | 0.2730490000 |
| H | -3.2794620000 | -1.4933500000 | -0.5519580000 |

D

| | | | |
|---|---|---|---|
| C | 1.7008540000 | -1.9923070000 | -0.6162310000 |
| C | 1.6396660000 | -1.1796140000 | 0.6686030000 |
| H | 0.7056500000 | -2.2855950000 | -0.9577790000 |
| H | 2.2744530000 | -2.9074780000 | -0.4552700000 |
| H | 2.1820740000 | -1.4303740000 | -1.4182820000 |
| H | 1.1215970000 | -1.7435210000 | 1.4496290000 |
| C | 0.9321070000 | 0.1722800000 | 0.5486510000 |
| H | 2.6506740000 | -0.9822160000 | 1.0381780000 |
| O | 1.7388400000 | 0.9342790000 | -0.3374190000 |
| H | 0.9240410000 | 0.6565450000 | 1.5340080000 |
| C | -0.4666170000 | 0.0469760000 | 0.0404700000 |
| C | -1.5535840000 | 0.1428450000 | 0.8047050000 |
| C | -2.9096150000 | -0.0376900000 | 0.2562350000 |
| O | -3.1567650000 | -0.3502960000 | -0.8834010000 |
| H | -0.5896570000 | -0.1525180000 | -1.0216090000 |
| H | -1.4774590000 | 0.3764310000 | 1.8641430000 |
| H | -3.7351560000 | 0.1389270000 | 0.9762470000 |
| O | 1.1682830000 | 2.2460050000 | -0.4256510000 |
| H | 1.8840430000 | 2.7749570000 | -0.0521030000 |

| E | | | | F | | | |
|---|---|---|---|---|---|---|---|
| C | -1.3526350000 | 2.0387340000 | -0.3643280000 | C | 2.9648410000 | -1.1647090000 | -0.2358960000 |
| C | -1.3805160000 | 1.0284220000 | 0.7716990000 | C | 1.5152760000 | -0.9354240000 | -0.6306310000 |
| H | -0.3323570000 | 2.3303710000 | -0.6243060000 | H | 3.4753730000 | -0.2175890000 | -0.0554490000 |
| H | -1.8924430000 | 2.9457610000 | -0.0846670000 | H | 3.4995130000 | -1.6930910000 | -1.0280610000 |
| H | -1.8213030000 | 1.6366780000 | -1.2659070000 | H | 3.0393900000 | -1.7652140000 | 0.6752780000 |
| H | -0.8483880000 | 1.4206030000 | 1.6438300000 | H | 1.4594590000 | -0.3612860000 | -1.5590510000 |
| C | -0.7654480000 | -0.3304830000 | 0.4292630000 | C | 0.7174820000 | -0.2025360000 | 0.4433020000 |
| H | -2.4090660000 | 0.8439540000 | 1.0945150000 | H | 1.0120990000 | -1.8890130000 | -0.8188520000 |
| O | -1.4398910000 | -0.9806270000 | -0.6470000000 | O | 1.3201840000 | 1.0342890000 | 0.8261280000 |
| H | -0.8165790000 | -0.9792380000 | 1.3113540000 | H | 0.7711230000 | -0.7627650000 | 1.3839250000 |
| C | 0.6478570000 | -0.2158170000 | -0.0354020000 | C | -0.7088850000 | -0.0074620000 | 0.0490840000 |
| C | 1.7119530000 | -0.4284220000 | 0.7373710000 | C | -1.7430700000 | -0.6439900000 | 0.5979940000 |
| C | 3.0845770000 | -0.2431020000 | 0.2298270000 | C | -3.1269330000 | -0.4135980000 | 0.1424660000 |
| O | 3.3613670000 | 0.1701540000 | -0.8700140000 | O | -3.4427090000 | 0.3457990000 | -0.7415080000 |
| H | 0.8088070000 | 0.0889570000 | -1.0671080000 | H | -0.9024240000 | 0.6946490000 | -0.7592970000 |
| H | 1.6062080000 | -0.7581940000 | 1.7678450000 | H | -1.6029250000 | -1.3559360000 | 1.4080540000 |
| H | 3.8900450000 | -0.5184960000 | 0.9408090000 | H | -3.9030540000 | -0.9965230000 | 0.6793460000 |
| O | -2.7413040000 | -1.3446490000 | -0.1741940000 | O | 1.4391110000 | 1.8615500000 | -0.3410440000 |
| H | -3.3010340000 | -0.8654220000 | -0.7972770000 | H | 0.9064890000 | 2.6199760000 | -0.0724180000 |
| G | | | | | | | |
| C | 1.6540730000 | -1.6353030000 | -0.9174200000 | | | | |
| C | 1.6073130000 | -1.1682290000 | 0.5303910000 | | | | |
| H | 0.6520370000 | -1.8249490000 | -1.3105530000 | | | | |
| H | 2.2133920000 | -2.5711840000 | -0.9865940000 | | | | |
| H | 2.1360410000 | -0.9010580000 | -1.5616630000 | | | | |
| H | 1.1581160000 | -1.9546760000 | 1.1468340000 | | | | |
| C | 0.8237120000 | 0.1147800000 | 0.8261750000 | | | | |
| H | 2.6199440000 | -1.0113810000 | 0.9152730000 | | | | |
| O | 1.5505040000 | 1.3113090000 | 0.5507330000 | | | | |
| H | 0.7211080000 | 0.1967430000 | 1.9139350000 | | | | |
| C | -0.5313800000 | 0.1412620000 | 0.1999690000 | | | | |
| C | -1.6677700000 | -0.1330720000 | 0.8401490000 | | | | |
| C | -2.9703210000 | -0.1102720000 | 0.1486680000 | | | | |
| O | -3.1238190000 | 0.1441090000 | -1.0216350000 | | | | |
| H | -0.5821360000 | 0.3841850000 | -0.8584960000 | | | | |
| H | -1.6766160000 | -0.3813820000 | 1.8989710000 | | | | |
| H | -3.8466450000 | -0.3462560000 | 0.7865340000 | | | | |
| O | 1.7913860000 | 1.4074100000 | -0.8614390000 | | | | |
| H | 1.3664340000 | 2.2523410000 | -1.0530990000 | | | | |

**S2. C$_6$-HPALD conformer energies and Boltzmann weights**

Energies of these geometries were then refined at df-CCSD(T)-F12/cc-pVDZ-f12 with def2-QZVPP basis set used for the density fitting, as in the calculations reported in Peeters *et al.*[4]. The energies calculated with df-CCSD(T) are those used to calculate conformer weights in the final run. The Boltzmann weight of conformer *i* with energy $E_i$ at 298K was calculated with the following formula (where $E_{min}$ is the energy of the most stable conformer).

$$w_i = \frac{e^{-(E_i - E_{min})/kT}}{\sum_i^{N_{conf}} e^{-(E_i - E_{min})/kT}}$$

The resulting weights are presented in this table (Energies from df-CCSD(T))

Table S1 Conformer weights calculated from energies computed with df-CCSD(T)-F12/cc-pVDZ-f12/def2-QZVPP

| Conformer | Energy (a.u.) | Weight | No of trajectories ran | No trajectories used in analysis |
|---|---|---|---|---|
| A | -459.6193492 | 0.224 | 50 | 24 |
| B | -459.6193492 | 0.459 | 50 | 50 |
| C | -459.6179138 | 0.049 | 10 | 5 |
| D | -459.6189484 | 0.146 | 20 | 16 |
| E | -459.61729 | 0.025 | 5 | 3 |
| F | -459.6185605 | 0.097 | 20 | 11 |
| G | -459.6148756 | 0.002 | 0 | 0 |

## S3. Electronic structure benchmarks for $C_6$-HPALD

Calculating properties of excited states, such as excited state energies, nuclear gradients, and nonadiabatic couplings (NAC) for larger molecules can be done accurately and efficiently with linear response (LR-)TDDFT. TDDFT is a single reference method and is formally exact. Its shortfalls are well documented, including its tendency to underestimate energies of states with high charge transfer character[5] or regions of the PES with strong coupling between ground and excited states[6]. Nevertheless, it is widely used for nonadiabatic dynamics simulations of larger systems due to its favourable computational cost.

Benchmarking employed a number of electronic structure methods which included DFT, LR-TDDFT, CIS, CIS(D), ADC(2), EOM-CCSD, SA-CASSCF, and MS-CASPT2. All DFT, LR-TDDFT, CIS, and CIS(D) calculations were performed in Gaussian 16[3]. CC2 and ADC(2) calculations were performed in Turbomole v7.1[7]. The EOM-CCSD calculation is executed in Molpro 2018[8]. CASSCF and MS-CASPT2 calculations were performed in OpenMolcas v18.09 [9]. The active space contained 10 electrons in 8 orbitals which includes σ/σ* orbitals at the peroxide bond, π/π* orbitals at the enone chromophore, and lone pairs on oxygen atoms.

All benchmarks were calculated for $n$=5 except CASSCF and MS-CASPT2 which were calculated for $n$=3.

Table S2 Excitation energies in eV for the first $n$ singlet states at the PBE0/TZVP optimised $S_0$ geometry of conformer B.

| Method | Basis Set | $S_1$ | $S_2$ | $S_3$ | $S_4$ | $S_5$ |
|---|---|---|---|---|---|---|
| CIS | 6-31G | 4.4425 | 6.7454 | 6.7915 | 8.3597 | 8.3927 |
| | 6-311+G* | 4.6675 | 6.4277 | 6.9813 | 7.9840 | 8.2861 |
| | 6-311++G** | 4.6457 | 6.4155 | 6.9665 | 7.7852 | 8.2444 |
| CIS(D) | 6-31G | 3.9083 | 6.6660 | 6.8104 | 7.7418 | 7.9101 |
| | 6-311+G* | 3.7471 | 6.1951 | 6.4346 | 6.9162 | 7.6128 |
| | 6-311++G** | 3.7468 | 6.1739 | 6.4064 | 6.9186 | 7.3769 |
| ωB97XD | 6-31G | 3.6468 | 5.7016 | 6.0060 | 6.1120 | 6.9943 |
| | 6-311+G* | 3.6694 | 5.6377 | 5.8160 | 6.1112 | 6.9448 |
| | 6-311++G** | 3.6461 | 5.6161 | 5.8004 | 6.0781 | 6.9032 |
| PBE0 | 6-31G | 3.5260 | 4.8622 | 5.8700 | 5.9635 | 6.2205 |
| | 6-311+G* | 3.5291 | 4.8701 | 5.6063 | 5.9754 | 6.1971 |
| | 6-311++G** | 3.5046 | 4.8478 | 5.5920 | 5.9338 | 6.1755 |
| LC-ωPBE | 6-31G | 3.6944 | 6.0125 | 6.2462 | 6.6052 | 7.4603 |
| | 6-311+G* | 3.7482 | 5.9197 | 6.1152 | 6.5375 | 7.4241 |
| | 6-311++G** | 3.7245 | 5.9069 | 6.0937 | 6.5056 | 7.2600 |
| CAM-B3LYP | 6-31G | 3.6584 | 5.7352 | 5.9908 | 6.1034 | 7.0161 |
| | 6-311+G* | 3.6840 | 5.6435 | 5.8159 | 6.0484 | 6.9394 |
| | 6-311++G** | 3.6599 | 5.6240 | 5.7978 | 6.0136 | 6.8136 |
| CC2 | cc-pVDZ | 3.8107 | 5.9433 | 6.4220 | 6.4997 | 7.4003 |
| ADC(2) | cc-pVDZ | 3.6180 | 5.9074 | 6.3592 | 6.4981 | 7.3464 |
| | cc-pVTZ | 3.5412 | 5.7588 | 6.1160 | 6.3720 | 7.1849 |
| | cc-pVQZ | 3.5151 | 5.7238 | 6.0326 | 6.2971 | 7.1033 |
| | aug-cc-pVDZ | 3.4504 | 5.5908 | 5.8035 | 5.9428 | 6.0525 |
| EOM-CCSD | aug-cc-pVDZ | 3.756 | 6.114 | 6.156 | 6.479 | 6.852 |
| SA(4)-CASSCF(8/10) | 6-31G* | 2.63 | 6.22 | 6.62 | - | - |
| MS(4)-CASPT2(8/10) | 6-31G* | 3.70 | 7.15 | 8.17 | - | - |

Rigid scans along the -O--OH coordinate of $C_6$-HPALD initiated at the optimised PBE0/TZVP $S_0$ geometry of conformer B.

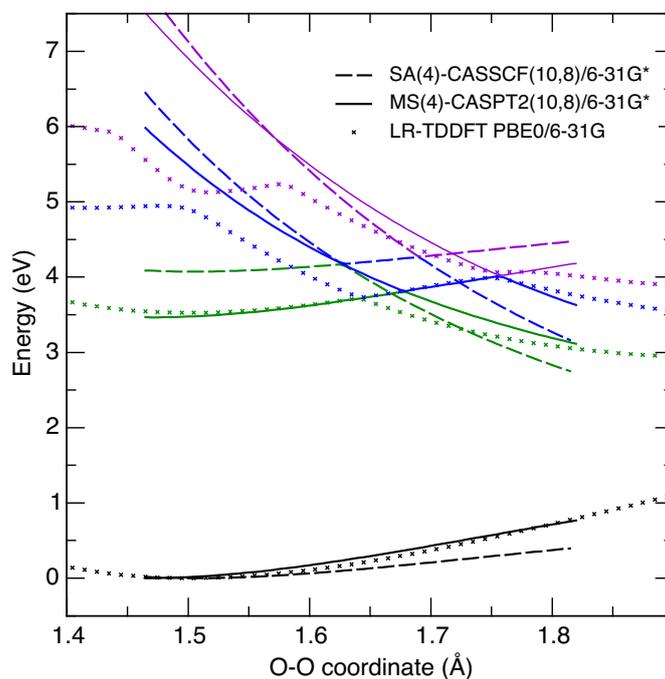

Figure S15 Benchmark scan showing profiles of the first 4 singlet states calculated with LR-TDDFT/PBE0/6-31G; CASSCF(10,8)/6-31G*; MS(4)-CASPT2(10,8)/6-31G*.

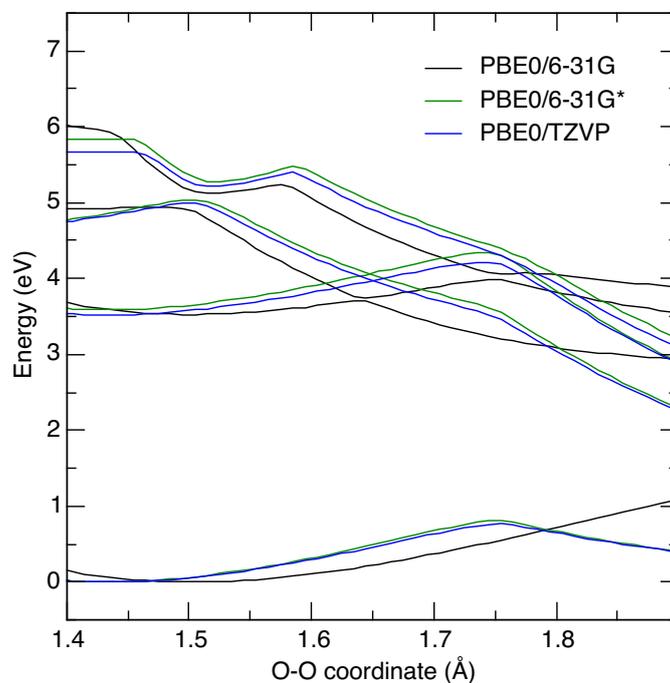

Figure S16 Benchmark scan showing profiles of the first 4 singlet states calculated with LR-TDDFT/PBE0 using basis sets 6-31G, 6-31G*, and TZVP.

## S4. Photoabsorption cross section of C$_6$-HPALD

The photoabsorption cross section σ(λ) is calculated using the Wigner ensemble sampling method described in Crespo-Otero et al.[10] which captures the inhomogeneous broadening of the spectral bands. Harmonic ground state frequencies used to generate the Wigner distribution were calculated for each conformer with DFT/PBE0/6-31G. Distribution assumes temperature of 0 K. For each of 7 conformers, 100 points are sampled from their respective distribution. We calculate the absorption in the 300-400nm range into the S$_1$ and S$_2$ electronic states separately, and the combined spectrum. For each sample point the vertical transitions and oscillator strengths are calculated with LR-TDDFT/PBE0/TZVP. Each peak is overlaid with a Lorentzian curve whose phenomenological broadening is set to 0.05 eV to return a continuous spectrum. The final photoabsorbtion cross section shown in Figure S3 is a linear combination of the spectra for each conformer $\sigma(\lambda) = \sum_i w_i \sigma_i(\lambda)$ where the weights $w_i$ are given in Table 1.

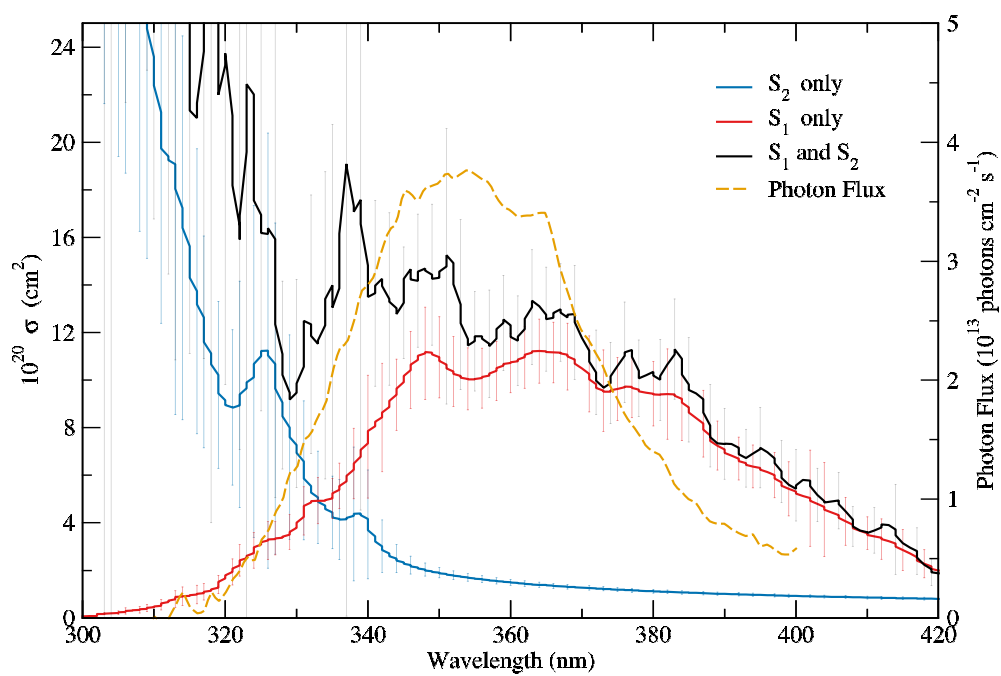

Figure S17 Calculated photoabsorbtion cross section of C$_6$-HPALD calculated using the Wigner ensemble method including excitations into the S$_1$ state only; the S$_2$ state only; both S$_1$ and S$_2$.

The photodissociation rate constant $J$ is then computed by integrating the product of the cross section σ, the incident flux $F$ and quantum yield $\phi$ (assumed to be 1). Values for the actinic flux are generously provided by John D. Crounse to replicate the experiments in Wolfe et al.[1].

$$J = \int_{\lambda_{min}}^{\lambda_{max}} \sigma(\lambda)\phi(\lambda)F(\lambda)d\lambda$$

## S5. Optimised geometries at the critical points on the S$_1$ surface:

| S$_1$ Transition State | | | | S$_1$/S$_2$ MECI | | | |
|---|---|---|---|---|---|---|---|
| C | -1.4060189 | 2.4098252 | 0.2235875 | C | 2.88515471 | -0.92818978 | -0.75884241 |
| C | -1.6798263 | 1.0453553 | -0.4014833 | C | 1.41203965 | -0.59288296 | -0.97515944 |
| H | -1.6007691 | 2.3975766 | 1.3024825 | H | 3.42940314 | -0.05369480 | -0.39234114 |
| H | -2.0497346 | 3.1755590 | -0.2214490 | H | 3.34760020 | -1.24905244 | -1.69791842 |
| H | -0.3649798 | 2.7180209 | 0.0745596 | H | 3.01043916 | -1.73688608 | -0.02837435 |
| H | -2.7220153 | 0.7451016 | -0.2513588 | H | 1.31457357 | 0.27216548 | -1.63921166 |

| | | |
|---|---|---|
| C  -0.8010352  -0.0566923   0.1959860 | C   0.66210618  -0.28007680   0.32891398 |
| H  -1.5027929   1.0714333  -1.4846103 | H   0.88880100  -1.43181368  -1.45325767 |
| O  -1.1589852  -1.3493592  -0.3627193 | O   1.35482785   0.73667542   1.10583675 |
| H  -0.9464474  -0.1069402   1.2850243 | H   0.71758141  -1.15563622   0.99648667 |
| C   0.6382591   0.0413216  -0.1641628 | C  -0.76116032   0.11248158   0.09942736 |
| C   1.6451761  -0.2963838   0.7070148 | C  -1.81437710  -0.72706045   0.33631882 |
| C   3.0178652  -0.2804158   0.3787539 | C  -3.16917057  -0.41435878   0.06581716 |
| O   3.5344058   0.0333306  -0.7568154 | O  -3.61173010   0.70295957  -0.40744809 |
| H   0.8700502   0.3361463  -1.1848627 | H  -0.91383261   1.10531042  -0.31477188 |
| H   1.4044025  -0.6021622   1.7207028 | H  -1.63942947  -1.71652301   0.75223380 |
| H   3.7456025  -0.5533040   1.1560210 | H  -3.93668791  -1.17476847   0.25961051 |
| O  -2.7049646  -1.6588353   0.1492353 | O   1.46384836   2.05283537   0.09424441 |
| H  -2.8058693  -2.4748824  -0.3875351 | H   1.82246888   2.64528928   0.79073068 |

**S6. Energies, rotational constants, and frequencies calculated at the critical points.**

All of the below values are calculated in Gaussian 16[3], with LR-TDDFT/PBE0/6-31G using tight convergence criteria and an ultrafine integration grid.

| | $S_1$ minimum (B) | $S_1$ minimum (C) | TS – $S_1$ | CI – $S_1/S_2$ [*] |
|---|---|---|---|---|
| Energy (a.u.) | -459.429938897 | -459.429169893 | -459.42369136 | -459.418027248 |
| ZPE correction (a.u) | 0.153697 | 0.153386 | 0.150686 | 0.1619025445 |
| Rotational constants | 0.06178 | 0.07586 | 0.07539 | 0.0222 |
| | 0.03267 | 0.02710 | 0.02689 | 0.0251 |
| | 0.02342 | 0.02170 | 0.02155 | 0.0938 |
| Frequencies | 53.7093 | 42.3725 | -3534.0736 | 68.62 |
| | 69.0555 | 68.6738 | 63.7551 | 78.221 |
| | 91.7755 | 93.1878 | 67.9529 | 84.668 |
| | 134.5593 | 112.3241 | 95.3035 | 128.213 |
| | 145.3678 | 147.5971 | 110.0446 | 181.793 |
| | 171.9759 | 164.3655 | 148.6376 | 223.116 |
| | 221.3219 | 210.9118 | 207.6926 | 253.395 |
| | 225.2273 | 233.587 | 228.0155 | 254.783 |
| | 299.9763 | 278.9438 | 253.1427 | 301.067 |
| | 352.2649 | 300.8378 | 275.4055 | 329.293 |
| | 387.0304 | 388.1045 | 308.6733 | 362.679 |
| | 390.6556 | 447.713 | 364.2645 | 410.98 |
| | 478.609 | 461.5256 | 434.6985 | 485.264 |
| | 517.1127 | 512.0716 | 456.206 | 583.909 |
| | 649.0519 | 660.1531 | 497.9537 | 698.771 |
| | 709.1161 | 719.9368 | 680.2171 | 771.322 |
| | 801.9211 | 789.8532 | 791.6283 | 793.207 |
| | 827.045 | 829.3933 | 826.7497 | 872.964 |
| | 892.4528 | 834.7936 | 865.5645 | 915.326 |
| | 906.2169 | 905.0289 | 905.0682 | 960.249 |
| | 943.5918 | 944.631 | 956.5599 | 1036.699 |
| | 955.5396 | 966.1619 | 959.6579 | 1058.41 |
| | 1026.4537 | 1031.0846 | 1025.5298 | 1087.157 |
| | 1058.011 | 1061.9474 | 1071.0154 | 1093.907 |
| | 1107.599 | 1106.8674 | 1085.7322 | 1137.154 |
| | 1180.0393 | 1172.6581 | 1110.114 | 1163.242 |
| | 1184.4502 | 1185.0541 | 1132.9102 | 1185.308 |
| | 1214.5041 | 1216.4378 | 1162.4788 | 1262.67 |
| | 1291.3523 | 1272.7574 | 1213.9031 | 1287.216 |
| | 1311.938 | 1296.4804 | 1274.7914 | 1310.993 |
| | 1324.6155 | 1327.2453 | 1292.0703 | 1338.057 |
| | 1342.4468 | 1343.5821 | 1315.029 | 1352.212 |
| | 1343.5943 | 1346.6629 | 1337.7966 | 1376.988 |
| | 1393.5145 | 1393.9353 | 1345.5383 | 1398.868 |
| | 1428.1916 | 1425.6609 | 1379.918 | 1455.576 |
| | 1461.3167 | 1461.0041 | 1414.6639 | 1521.33 |
| | 1499.3381 | 1502.7229 | 1461.0081 | 1539.105 |
| | 1539.3167 | 1540.7134 | 1539.1839 | 1549.307 |

| | | | |
|---|---|---|---|
| 1543.0902 | 1543.8922 | 1542.1446 | 1603.235 |
| 1550.8704 | 1550.9794 | 1549.4011 | 3018.911 |
| 1598.7604 | 1609.256 | 1604.491 | 3067.174 |
| 3066.7762 | 3062.8107 | 3049.8769 | 3067.697 |
| 3075.598 | 3064.5537 | 3050.92 | 3070.362 |
| 3079.3257 | 3072.6858 | 3063.5567 | 3134.348 |
| 3091.1291 | 3075.6386 | 3078.6837 | 3153.494 |
| 3129.6292 | 3129.3525 | 3134.677 | 3177.919 |
| 3156.2406 | 3152.4157 | 3154.9103 | 3209.011 |
| 3162.2071 | 3158.2418 | 3159.2982 | 3235.128 |
| 3211.1928 | 3222.2468 | 3218.3714 | 3644.865 |
| 3237.3168 | 3238.926 | 3232.116 | |
| 3623.5612 | 3650.8246 | 3646.0593 | |

## S7. Branching space at the MECI

| Gradient difference (g) | | | | Derivative coupling (ETF corrected) (h) | | |
|---|---|---|---|---|---|---|
| C | 0.102056E-03 | -0.621470E-04 | -0.390281E-03 | C | -4.484668 | -4.838438 | 18.284332 |
| C | 0.297746E-03 | 0.211320E-03 | 0.118439E-02 | C | -12.150318 | 39.701432 | -69.764144 |
| H | -0.429340E-04 | 0.161000E-04 | 0.514220E-04 | H | 3.540401 | 0.003348 | -4.296759 |
| H | 0.156898E-03 | -0.630570E-04 | 0.194940E-04 | H | -7.843623 | 3.484781 | 1.387522 |
| H | -0.128200E-04 | 0.749900E-05 | 0.466640E-04 | H | -0.890071 | 0.491148 | -2.669058 |
| H | -0.926970E-04 | -0.730590E-04 | -0.737730E-04 | H | 7.949269 | 5.579241 | -2.076216 |
| C | -0.271767E-02 | -0.617192E-03 | -0.439607E-02 | C | 191.653690 | -145.164158 | 287.966800 |
| H | 0.179710E-04 | 0.226490E-04 | -0.988220E-04 | H | -2.392614 | -3.880188 | 9.528667 |
| O | 0.384463E-02 | 0.137730E-01 | -0.155984E-01 | O | -240.341234 | -1192.585893 | 767.883422 |
| H | 0.546384E-03 | 0.515009E-03 | 0.576439E-03 | H | -60.385774 | -51.681452 | -68.595717 |
| C | 0.349772E-02 | 0.206840E-02 | -0.369046E-03 | C | -279.340661 | -176.445168 | 48.087380 |
| C | -0.842186E-02 | -0.983095E-03 | -0.356232E-03 | C | 643.550588 | 81.899861 | 34.330309 |
| C | 0.106874E-01 | -0.965713E-02 | 0.992957E-02 | C | -841.867843 | 916.299623 | -461.119380 |
| O | -0.497100E-02 | 0.736638E-02 | -0.693473E-02 | O | 389.370823 | -688.428675 | 313.400606 |
| H | 0.553938E-03 | -0.484603E-03 | -0.168809E-03 | H | -33.073840 | 18.498041 | -1.989105 |
| H | -0.623553E-03 | -0.216590E-04 | -0.326236E-03 | H | 47.733292 | -4.781142 | 15.271671 |
| H | -0.874117E-03 | 0.883858E-03 | -0.855460E-03 | H | 68.546600 | -81.100604 | 38.675828 |
| O | -0.165508E-02 | -0.128428E-01 | 0.176537E-01 | O | 107.048607 | 1265.057512 | -913.533304 |
| H | -0.293047E-03 | -0.595600E-04 | 0.106157E-03 | H | 23.377372 | 17.890744 | -10.772851 |

The derivative coupling vector **h** was calculated at the MECI, (TDDFT/PBE0/6-31G), in Q-Chem, following the procedure described in Ou et al.[11]

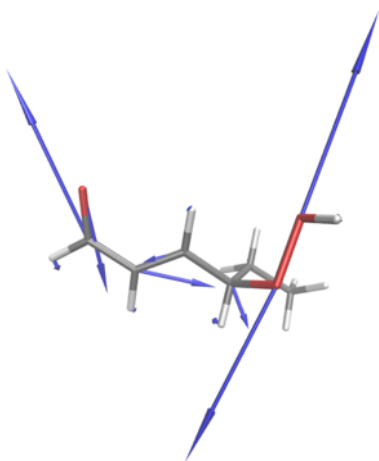

Figure S18 Derivative coupling vector at the $S_1/S_2$ MECI

## S8. Peroxide bond length coordinate R(O-O) throughout the NAMD simulation.

Figure S6 illustrates that for most cases trajectories that cross the 1.75 Å threshold will dissociate soon after. The average -O-OH bond length stays close to 1.5 Å throughout.

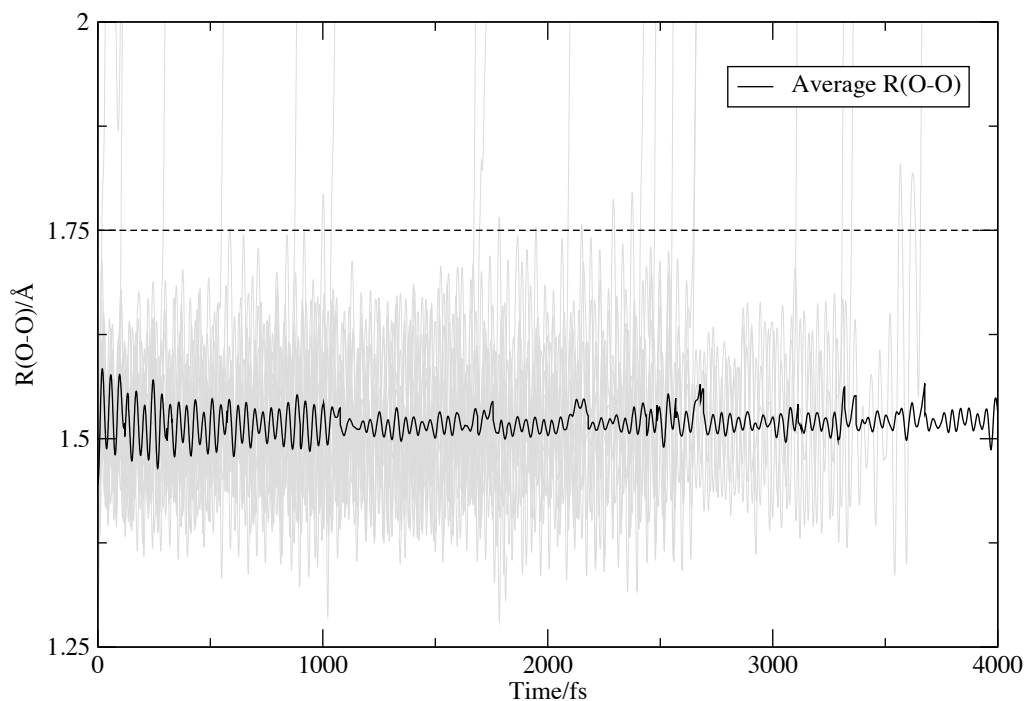

Figure S19 Peroxide bond length for each of 50 trajectories (initiated at the conformer B geometry) showing that crossing mostly occurred once the trajectory surpassed the 1.75 Å threshold.

**S9. Testing timestep sensitivity of FSSH**

Running FSSH requires a judicious choice of many parameters, one of which is the time step (dt) with which the classical degrees of freedom evolve on the PESs. In most cases, this decision is made by selecting the smallest dt for a given computational cost, to maximise the accuracy of the dynamics. However, what exactly is compromised when dt is too large?

For example, the accuracy of the classical integration of nuclear coordinates may be diminished. The more subtle point is that the transition probabilities between states also depend on time discretization[12]. At each time step the time derivative coupling (TDC) is calculated to determine if a hop to another adiabatic state occurs and when TDC is very localised it can be poorly resolved or missed altogether.

Meek et al. highlighted this problem for trivially unavoided crossings (TUC)[13] which are (3N-7) dimensional intersections between two weakly coupled states, where N is the number of nuclear degrees of freedom. If the coupling between the states is infinitesimal, a crossing trajectory must necessarily remain on the same diabatic state. As it passes through the seam there would be a sudden narrow spike in the TDC which might not be resolved with a larger dt, and so the trajectory will, incorrectly, move to a different diabatic state. For (3N-8) dimensional conical intersections this problem is less significant, as the nonadiabatic coupling is highly localised in position space.

In this case, the topology of the $C_6$-HPALD seam can be described as a TUC as there is an effective (3N-7) crossing at the intersection of the $n\pi^*$ and $n'\sigma^*$ diabatic states with a weak nonadiabatic coupling between them. Slowness of the dissociation process limits us in how small the time step can be, but it becomes necessary to verify the extent to which this might affect the outcomes of the FSSH dynamics.

Although the loss of OH typically takes picoseconds, many trajectories encounter the (3N-7) seam in the first 50 fs of the simulation. It is then possible to explore how the shape of the TDC is affected by the timestep and to see if the spike in TDC is too sharp to detect with dt=0.5 fs. Three example trajectories were taken from the set initiated at conformer C and repeated for 50 fs using identical initial conditions for dt = 0.25; dt = 0.1 fs; dt = 0.05 fs; dt = 0.025 fs. Results are shown in Figure S20.

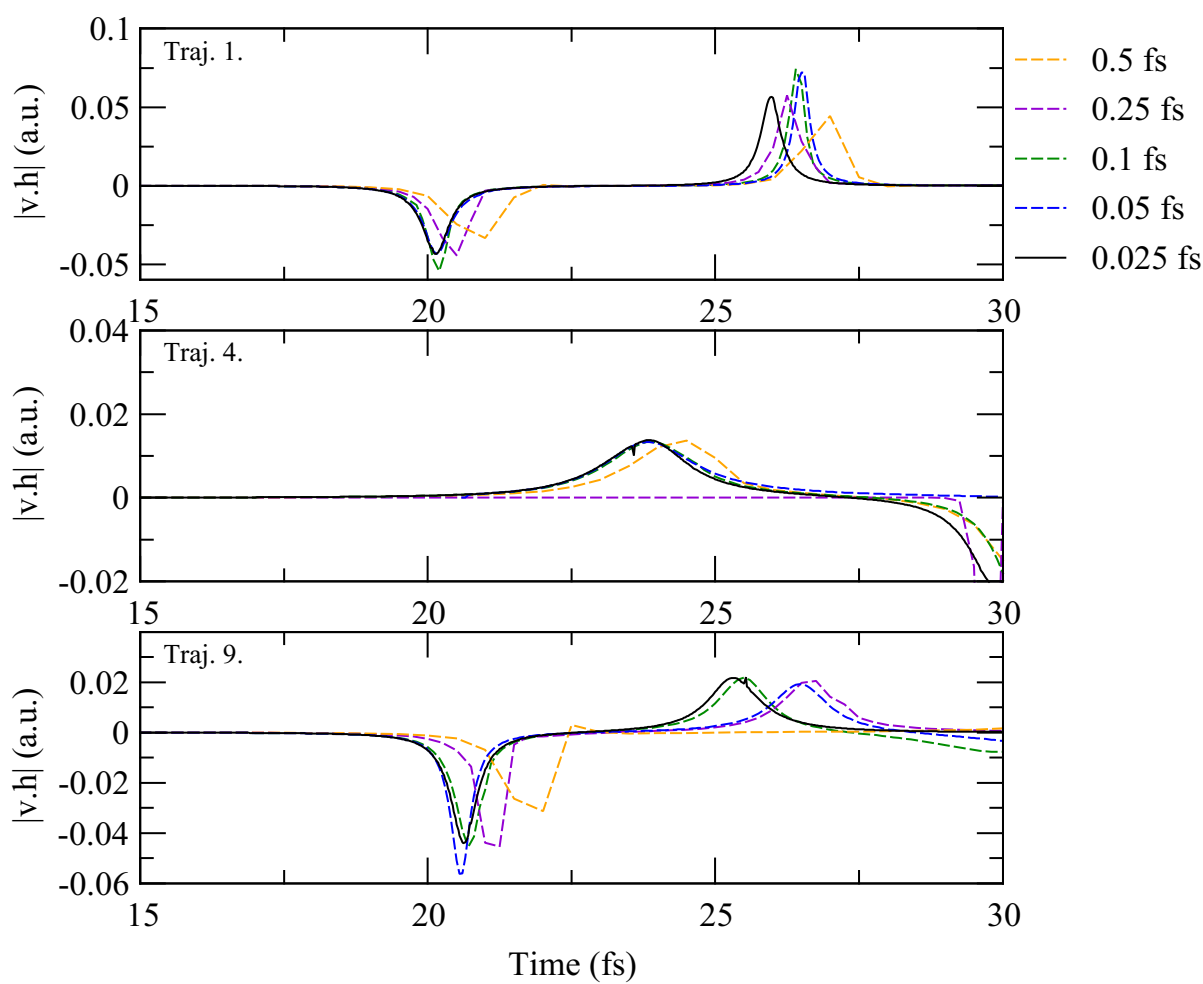

Figure S20 Convergence of time derivative coupling with time step size for three example trajectories in the first 30 fs of the simulation.

Here we see the results converging with progressively smaller values of dt. Ideally, the trajectories should be run with at least a 0.1 fs timestep to properly capture the shape of the TDC, however the computational cost of doing this is prohibitive. The peaks are still present for 0.5 fs time step trajectories, even though they are visibly coarser.

Most trajectories encounter this seam multiple times, so it is still possible that the error in the shape of the TDC peak for 0.5 fs time step may affect the long-time dynamics of OH loss in HPALD. Full length trajectories were run to look for trends in dissociation behaviour between dt = 0.25 fs and dt = 0.5 fs. For each of a set of 3 IC, we ran 20 replicas which were seeded with a different random number. Results can be seen in Figure S21 . These results appear to indicate that the time step size makes no clear difference to the rate of OH loss and so on this basis we conclude that it is safe to use a 0.5 fs time step for FSSH dynamics.

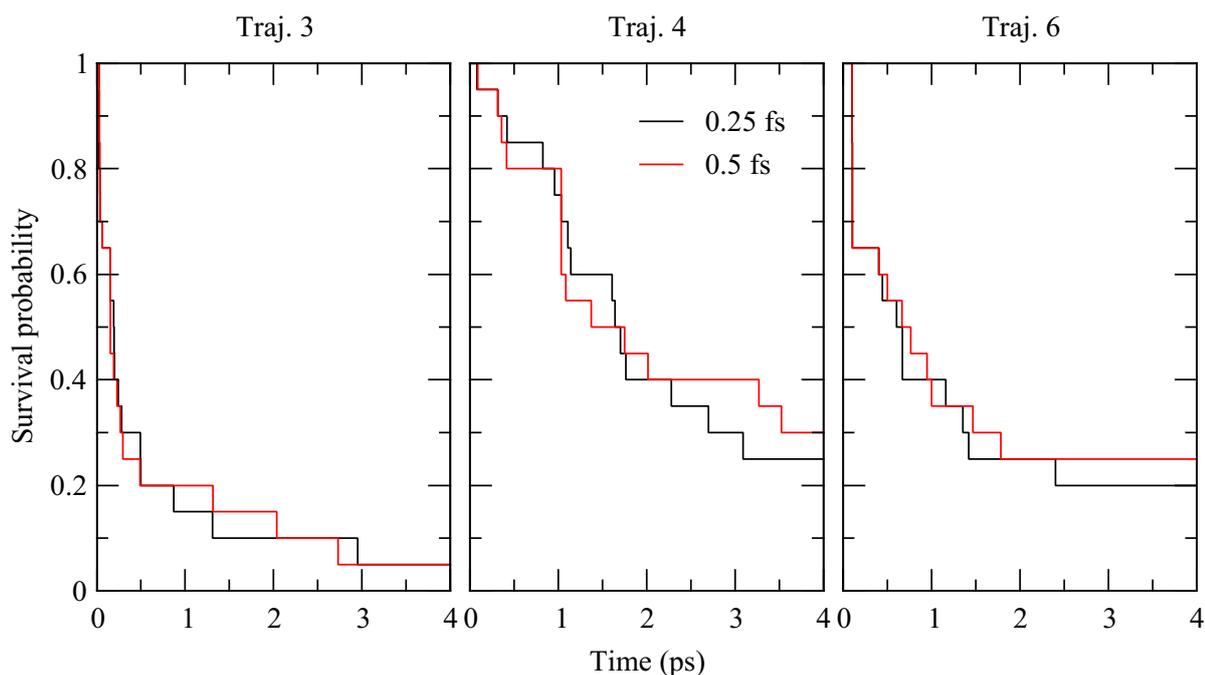

Figure S21 Impact of time step size on the rate of OH loss in HPALD.

**S10. Parameters of the diabatic states fitted to the adiabatic surface scan across the eigenvector coordinate (with LR-TDDFT/PBE0)**

Implementation of the Zhu-Nakamura equation in MESMER requires an analytic form of the diabatic crossing. These equations were fitted to the
Bound diabatic potential $E_{\pi^*}(R)$ fitted to a harmonic function
$$E_{\pi^*}(R) = A_{\pi^*} \times (R - \beta_{\pi^*})^2 + \varepsilon_{\pi^*}$$
Dissociative diabatic potential $E_{\sigma^*}(R)$ fitted to an exponential decay
$$E_{\sigma^*}(R) = A_{\sigma^*} \times e^{-R \times \beta_{\sigma^*}} + \varepsilon_{\sigma^*}$$

Parameter fitting was performed in Gnuplot v5.2 with resulting parameters are listed in the following table:

| Parameter | Value | Standard Error |
|---|---:|---:|
| $A_{\sigma^*}$ | 495.051 | ± 15.01 |
| $\beta_{\sigma^*}$ | -0.119809 | ± 0.00349 |
| $\varepsilon_{\sigma^*}$ | 11.0345 | ± 0.1939 |
| $A_{\pi^*}$ | 66.9186 | ± 2.388 |
| $\beta_{\pi^*}$ | 3.76346 | ± 1.307 |
| $\varepsilon_{\pi^*}$ | -46.4714 | ± 2.382 |
| $H_{12}$ | 3.00481 | ± 0.01141 |

**S11. Sensitivity testing the EGME model**

Some parameters included in the EGME calculation might have a significant impact on the result. We test the robustness of the EGME model by comparing the result when those parameters are varied.

An assumption we make in the EGME model is that it is valid to use the average initial energy from TSH initial conditions, rather than a distribution. Figure S9 shows results of an EGME model built to replicate the original 10 trajectories starting from the minimum of conformer C. A ZN-EGME calculation is performed for each trajectory IC illustrating that the model is robust to modest variation in initial energy.

EGME results can also be sensitive to frequencies, especially low frequencies that have a significant impact on the density of states. Figure S9 compares EGME results calculated using critical points optimised with tight convergence conditions and a fine integration grid against results calculated with a standard grid and convergence criteria. Even with a slight difference in frequencies the decay curves are visibly different. This highlights the importance of using frequencies calculated in a consistent way for each isomer when constructing an EGME model.

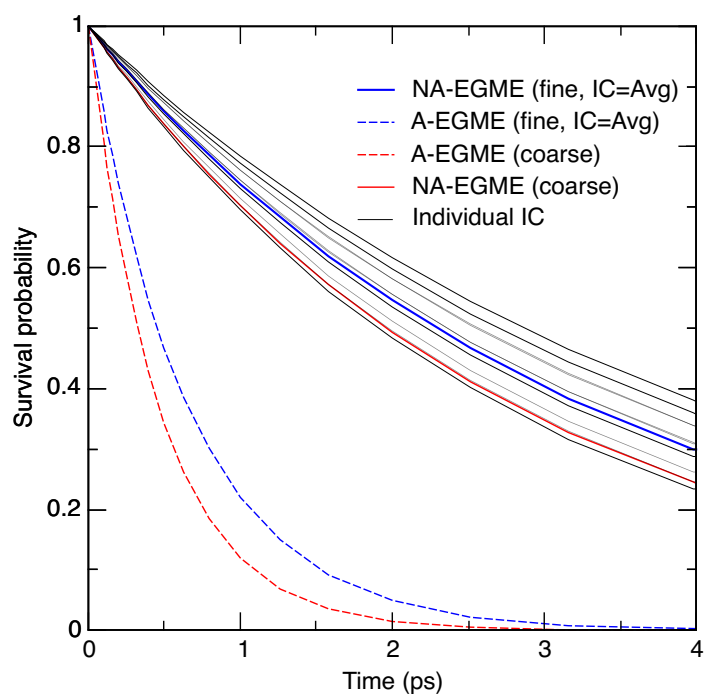

Figure S22 Result of EGME calculation intended to replicate the dissociation of the first 10 trajectories of conformer C.

## S12. Parameters of the diabatic states fitted to the adiabatic surface scan across the eigenvector coordinate (with MS(5)-CASPT2(10,8)/6-31G*)

Table S3 Fitted parameters for the diabatic potentials around the TS calculated with CASPT2.

| Parameter | Value | Standard Error |
|---|---|---|
| $A_{\sigma^*}$ | 835.318 | ± 73.88 |
| $\beta_{\sigma^*}$ | -0.113482 | ± 0.009752 |
| $\varepsilon_{\sigma^*}$ | 9.51291 | ± 0.8865 |
| $A_{\pi^*}$ | 50.3379 | ± 4.159 |
| $\beta_{\pi^*}$ | 5.85351 | ± 0.4645 |
| $\varepsilon_{\pi^*}$ | -29.9859 | ± 4.144 |
| $H_{12}$ | 0.823152 | ± 0.04453 |

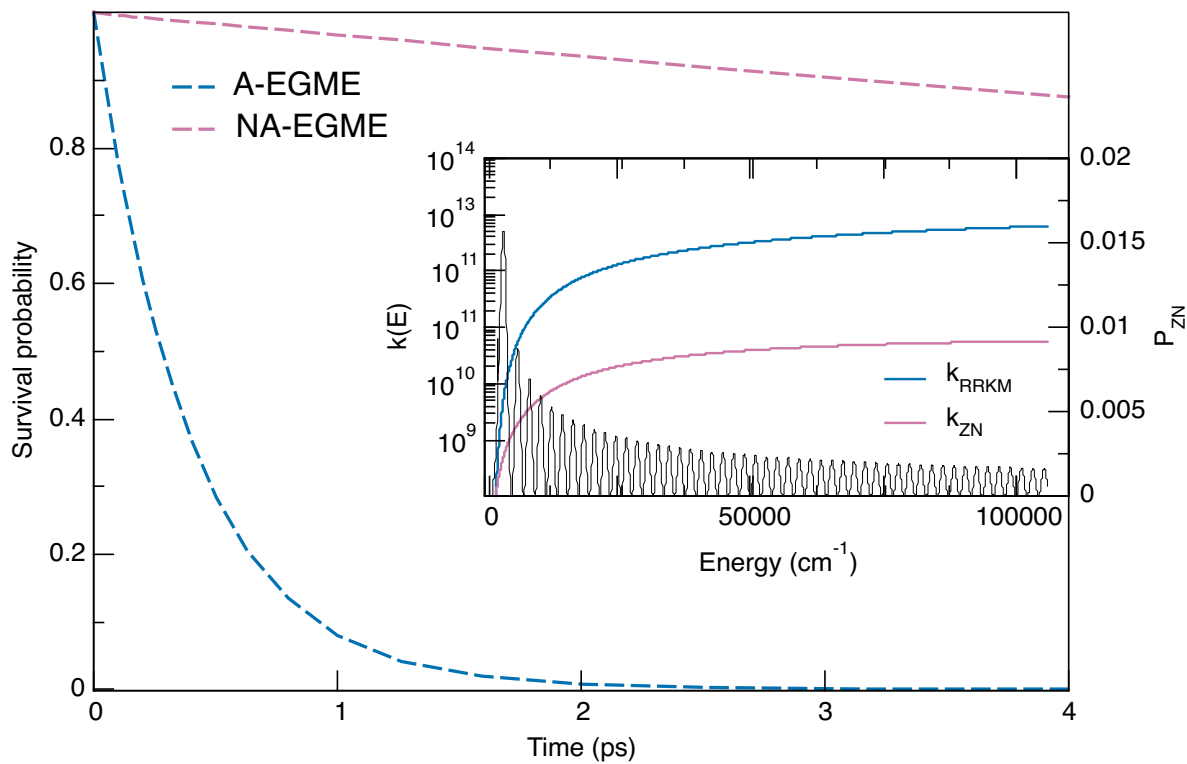

Figure S23 Results of EGME calculations based on MS-CASPT2(10,8) energies at the critical points. Inset contains the microcanonical rate coefficients calculated for each energy grain from RRKM (GS-EGME) and ZN (NA-EGME).

**S13. NA-EGME results with hindered rotor corrections.**

Results obtained by re-running the EGME calculations with the hindered rotor vibrations projected out of the Hessian. Using $S_1(B)$ minimum and TS, both optimised with tight convergence criteria and ultrafine integration grid.

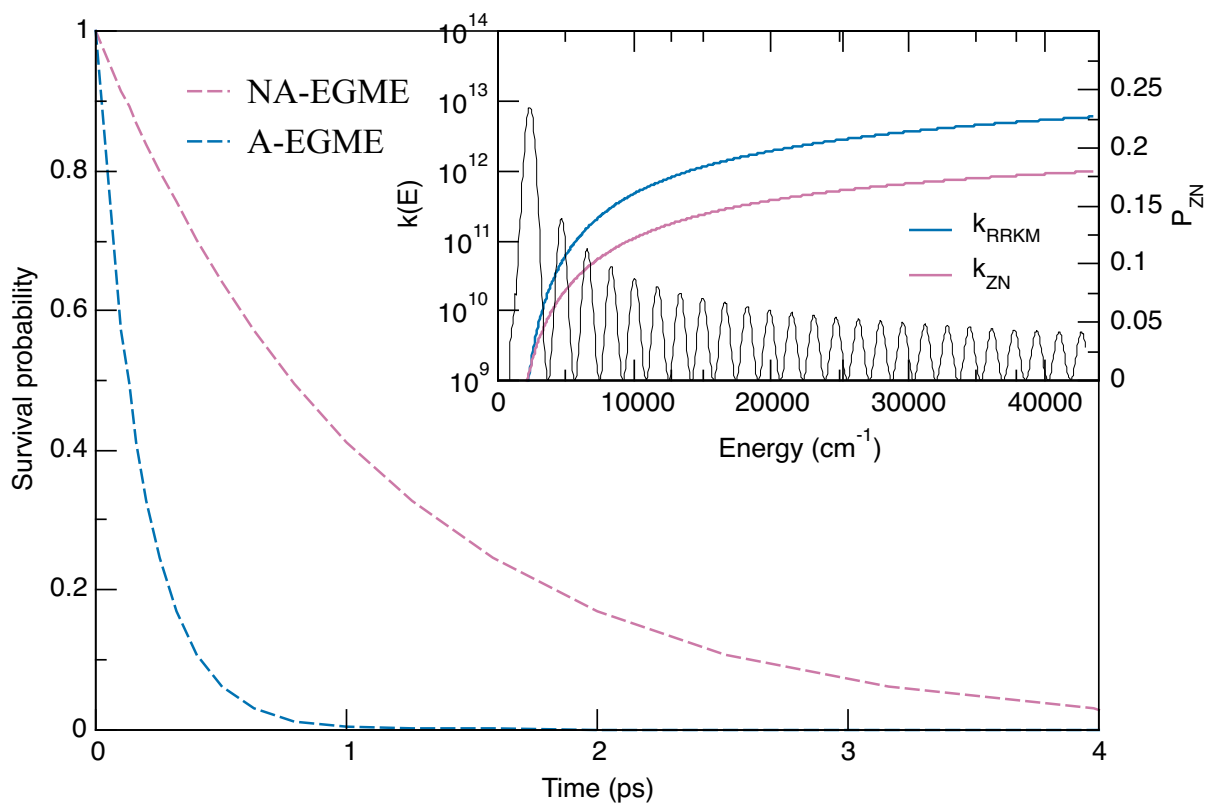

Figure S24 EGME results with hindered rotor corrections.

## S15. Zhu-Nakamura equations

This section contains the full set of Zhu-Nakamura equations, as implemented in MESMER[14].

Figure S25 is an example schematic of a nonadiabatic tunnelling (NT) type crossing, defined by the opposing slopes of the two crossing diabats. This is in contrast to a Landau Zener type crossing for which there is a similar set of equations, references where they can be found which are available. The figure highlights the features of the crossing that will be used in the following equations, both in the adiabatic (black) and diabatic (red) representations.

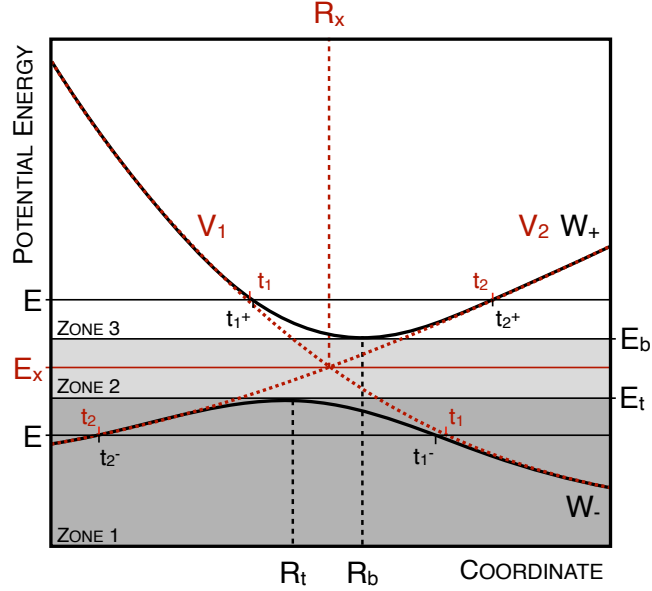

Figure S25: NT type crossing. Red, black features correspond to diabatic, adiabatic representations respectively.

Variable naming in the following equations will follow the conventions of the original paper. To improve comprehension of more cumbersome equations, dummy variables will be also be used and will be labelled $g_n$.

In the diabatic representation we define parameters $a^2$ and $b^2$ as
$$a^2 = \frac{\hbar^2 F(F_1 - F_2)}{16\mu V_X^3}$$
$$b^2 = (E - E_X) \times \frac{(F_1 - F_2)}{2FV_X}$$
where $F_n$ is the slope of the diabatic potential, $F = \sqrt{F_1 F_2}$, reduced mass of the system is $\mu$ and $V_X$ is the inter-state coupling. The crossing is split into 3 different energy ranges (zones – shown in Figure S25) which require separate sets of equations. In MESMER, the algorithm will solve for the Zhu-Nakamura transition probability $P_{12}$ at each energy grain.

Zone 1: $E \leq E_t$
We begin by calculating the integral of the tunnelling action $\delta$ through a single potential barrier on the adiabatic surface $W_-$
$$K_-(R) = \frac{1}{\hbar}\sqrt{2\mu(E - W_-(R))}$$
$$\delta = \int_{t_2^-}^{t_1^-} |K_-(R)| dR$$

Next, we need adiabatic parameter $\sigma$ which captures the effect of nonadiabatic transition between the two adiabatic surfaces. $\Gamma(X)$ is the standard Gamma function.
$$\sigma = \frac{\pi}{2a|b|} \frac{\left(6 + 10\sqrt{1 - \frac{1}{b^4}}\right)^{\frac{1}{2}}}{1 + \sqrt{1 - \frac{1}{b^4}}}$$

$$\sigma_c = \sigma\left(1 - 0.32 \times 10^{-\frac{2}{a^2}} e^{-\delta}\right)$$

$$B(X) = \frac{2\pi X^{2X} e^{-2X}}{X\Gamma^2(X)}$$

Broadly, $e^{-2\delta}$ can be interpreted as a Gamow factor for tunnelling and $B(\sigma_c/\pi)$ as the impact of nonadiabatic coupling on the tunnelling. Combining these factor yields the expression for the transition probability for energies $\leq E_t$

$$P_{12} = \frac{B(\sigma_c/\pi)e^{-2\delta}}{\left(\left(1 + \frac{a}{2(1+a)}B(\sigma_c/\pi)e^{-2\delta}\right)^2 + B(\sigma_c/\pi)e^{-2\delta}\right)}$$

Zone 2: $E_t \leq E \leq E_b$

We begin by defining two empirical corrections $g_1$ and $g_2$ that ensure the full range of coupling strengths is covered. $W$ is a factor of the Stokes constant.

$$g_1 = \frac{a - 3b^2}{a + 3}\sqrt{1.23 + b^2}$$

$$g_2 = 0.38\frac{(1 + b^2)^{1.2 - 0.4b^2}}{a^2}$$

$$W = \int_0^\infty \frac{1 + g_2}{a^{2/3}} \cos\left(\frac{t^3}{3} - \frac{b^2 t}{a^{\frac{2}{3}}} - \frac{g_1 t}{a^{\frac{2}{3}}(0.61\sqrt{2 + b^2} + a^{1/3}t)}\right) dt$$

Combining these factors gives us the transition probability.

$$P_{12} = \frac{W^2}{1 + W^2}$$

Zone 3: $E \leq E_b$

Integral of the classical action on the upper adiabatic surface $\sigma$ and nonadiabatic transition parameter $\delta$ are used to calculate the phase $\phi$

$$K_+(R) = \frac{1}{\hbar}\sqrt{2\mu(E - W_+(R))}$$

$$\sigma = \int_{t_1^+}^{t_2^+} |K_+(R)| dR$$

$$\delta = \frac{\pi}{16ab} \frac{\left(6 + 10\sqrt{1 - 1/b^4}\right)^{1/2}}{1 + \sqrt{1 - 1/b^4}}$$

$$\phi = \sigma + \frac{\delta}{\pi} - \frac{\delta}{\pi}\ln\left(\frac{\delta}{\pi}\right) + \frac{\pi}{4} - g_3 + \text{Arg}\,\Gamma\left(i\frac{\delta}{\pi}\right)$$

where $g_3$ is

$$g_3 = \frac{0.23a^{1/2}}{a^{1/2} + 0.75} 40^{-\sigma}$$

Finally, we use the Landau-Zener transition probability $p$

$$p = \exp\left(\frac{-\pi}{4a}\sqrt{\frac{2}{b^2 + (b^4 + g_4)^{1/2}}}\right)$$

where $g_4$ is

$$g_4 = -0.72 + 0.62a^{1.43}$$

to yield the final expression for the nonadiabatic transition probability in zone 3.

$$P_{12} = \frac{4\cos^2\phi}{4\cos^2\phi + p^2/(1-p)}$$

**S16. MESMER input files for:**

a) RRKM calculation of seam crossing rate – in file *rrkm.xml*
b) ZN calculation of seam crossing rate – in file *zn.xml*